\documentclass[letterpaper,twocolumn,10pt]{article}
\usepackage{usenix}   % loads: mathptmx, fontenc/inputenc, microtype, cite, url,
\usepackage{graphicx}
\usepackage{xspace}
\usepackage{booktabs}
\usepackage{multirow}
\usepackage{subcaption}
\usepackage{enumitem}
\usepackage{amsmath}
\usepackage{amssymb}
\usepackage{amsthm}
\usepackage{algorithm}
\usepackage{algorithmic}
\usepackage{placeins}
\usepackage{tcolorbox}
\tcbuselibrary{skins,breakable}

\newtheorem{proposition}{Proposition}

\theoremstyle{definition}

\definecolor{skillslate}{RGB}{62,95,125}
\definecolor{skillslatebg}{RGB}{242,246,250}
\newtcolorbox{rqbox}[1][]{
  breakable,
  enhanced,
  colback=white,
  colframe=white,
  frame hidden,
  borderline west={2pt}{0pt}{skillslate},
  left=8pt, right=4pt, top=2pt, bottom=2pt,
  before upper={\textbf{\textcolor{skillslate}{#1.}}\hspace{0.5em}},
}

\newcommand{\sysname}{\textsc{SkillShield}\xspace}

\newcommand{\etal}{\textit{et al.}\xspace}

\begin{document}
%-------------------------------------------------------------------------------

%don't want date printed
\date{}

\title{\Large \bf \sysname: Prompt-Space Security Skills for LLM Coding Agents}

\author{
  {\rm Xiaodong Wu\textsuperscript{*}, \quad Zhimin Zhao\textsuperscript{*}, \quad Qi Li, \quad Xiangman Li,}\\[2pt]
  {\rm Yu Shi, \quad Bram Adams, \quad Jianbing Ni}\\[3pt]
  {\rm Queen's University}\\[2pt]
  {\rm \{xiaodong.wu,\,z.zhao,\,qi.li,\,xiangman.li,\,y.shi,\,bram.adams,\,jianbing.ni\}@queensu.ca}\\[4pt]
  {\rm \textsuperscript{*}These authors contributed equally.}
}

\maketitle

\begin{abstract}
A coding agent edits files and executes shell commands with its developer’s privileges, allowing malicious requests to translate directly into harmful actions or functional malware. Existing defenses have complementary limitations: weight-level alignment is unavailable to API-only deployers, whereas input filters and execution-boundary monitors require auxiliary classification or checking components along the agent’s trajectory. We therefore introduce \sysname, a system-prompt defense that synthesizes \emph{security skills} offline from known attacks or recorded agent failures. These skills are injected into the system prompt at session start and remain active throughout the tool-use loop. Unlike a reference monitor, they protect the system by defining the security policies the model should follow during execution. Due to the limited system-prompt space, we examine three fixed-budget provisioning scopes: \emph{all-classes}, with one skill covering all threat classes, \emph{per-bundle}, with one skill targeting a related subset, and \emph{per-class}, with one skill dedicated to a single known class and used as the upper-bound reference. None requires runtime request classification or routing. Across six large language models on RedCode, the default \emph{all-classes} skill reduces malware-generation severity from 3.37 to 0.58 and achieves a 43.6\% execution attack success rate, comparable to Llama Guard~3’s 42.7\% without its separate 8B classifier. The \emph{per-bundle} and class-fixed \emph{per-class} settings further reduce this rate to 36.2\% and 14.5\%, respectively. Under two non-adaptive jailbreak families, \sysname continues to outperform all baselines on malware generation. Across 731 benign task descriptions, \sysname yields a mean safety-refusal rate of 0.14\%. These results demonstrate the potential of prompt-space security skills to prevent harmful actions and malware generation for LLM coding agents.
\end{abstract}

\section{Introduction}
\label{sec:intro}
% ============================================================

Large language model (LLM)-based coding agents such as Claude Code~\cite{claudecode2025}, Codex~\cite{openaicodex2021}, and OpenCode~\cite{opencode2026} combine language models with tools, code execution, and operating-system access. Conventional code assistants such as GitHub Copilot~\cite{githubcopilot2021} suggest code for human approval. Coding agents can also inspect repositories, edit files, invoke compilers and tests, execute shell commands, and debug failures iteratively~\cite{yang2024sweagent,wang2025openhands}. They now operate in development environments that contain credentials, dependency managers, build systems, and version-control state~\cite{hou2024large}.

Coding agents perform these actions with the developer's privileges and also process untrusted project content~\cite{greshake2023indirect,zhan2024injecagent,ruan2024toolemu}. A malicious GitHub issue can therefore instruct an agent diagnosing a build failure to fetch and execute a remote script, and the agent may comply~\cite{guo2024redcode,andriushchenko2024agentharm}. Because agent-generated tool calls may be executed without human approval, a harmful action can reach the execution boundary before a human intervenes. Runtime monitors can still block such candidate actions, but doing so requires inspecting each one along the trajectory. We instead study an earlier control point: placing security policy in the context that generates every action, so the model can be steered before a harmful call is produced. The need is substantial. Across six LLMs, undefended agents comply with 46--80\% of malicious code-execution requests, and DeepSeek-V3.2~\cite{deepseekai2025deepseekv32} also produces malware with an average judge score of 7.76 out of 10~\cite{gu2024survey}.

Existing defenses address this risk at four stages of the agent pipeline (Figure~\ref{fig:overview}). At the \emph{model level}, alignment~\cite{ouyang2022training,bai2022training,rafailov2024dpo} requires access to model weights and costly retraining, which excludes API-only deployers. At the \emph{input stage}, safety filters~\cite{metallamaguard3,metapromptguard2025} can reject harmful user requests, but they cannot protect against malicious instructions encountered later during the agent's interaction with repository content or tool output. At the \emph{execution stage}, runtime verifiers~\cite{luo2025agrail,jia2025task} inspect candidate actions before execution and provide a stronger enforcement boundary, but they add a model or checking component to the tool-use loop. In \emph{prompt space}, defenses place security instructions in the agent context. Existing methods use boundary reminders, provenance delimiters, or authentication tags to distinguish untrusted data from instructions~\cite{yi2025benchmarking,hines2024defending,wang2024fath}. These methods use general rules across tasks and leave open how threat-specific policies should be synthesized and scoped within a finite prompt budget. We investigate this question for API-only deployments. The system prompt can govern action formation throughout the multi-step loop and be updated without retraining or an additional runtime component.

%We introduce \sysname, a prompt-space defense that synthesizes \emph{security skills} and places them in a coding agent's system prompt. A skill is a natural-language instruction document based on the emerging Agent Skills standard~\cite{agentskills2025}. \sysname can synthesize a skill from known attacks through \emph{proactive mining} or from recorded agent failures through \emph{reactive learning}. The resulting skill remains in the system prompt throughout the session. Operators also determine its coverage before deployment. A general-purpose agent uses one \emph{all-classes} skill. An environment exposed to a known family of mechanisms can use a \emph{per-bundle} skill. A narrowly constrained environment with one known threat class can use a \emph{per-class} skill. All configurations use the same prompt budget and require no request-time classification or routing.

We therefore propose \sysname, a prompt-space defense that injects \emph{security skills} into a coding agent's system prompt. A skill is a natural-language instruction document that extends agent behavior without modifying model weights~\cite{agentskills2025}. \sysname derives security skills from structured attack data. \emph{Proactive mining} extracts observed attack patterns, whereas \emph{reactive learning} captures the framings behind recorded agent failures. Each skill is injected in full at session start and remains active throughout the interaction. Because the prompt budget is fixed, increasing coverage forces more threat classes to share the same space, reducing the detail available to each class, while narrowing scope preserves class-specific detail at the cost of covering fewer threats. To capture this tradeoff, we study three deployment scopes that reflect how much threat information is known before a session begins. General-purpose agents use an \emph{all-classes} skill, deployments exposing a known mechanism family use a corresponding \emph{per-bundle} skill, and \emph{per-class} skills apply when the exact threat class is known in advance and provide an upper bound on fine-grained provisioning. All configurations use the same prompt budget and require no request-time classification or routing.

We evaluate \sysname on RedCode~\cite{guo2024redcode}, the only publicly available and established coding-agent benchmark that jointly measures harmful tool execution and malware generation. Four questions organize the evaluation. \textbf{Q1: Are prompt-space security skills effective?} Yes. The all-classes skill reduces execution attack success rate (ASR) from 67.4\% to 43.6\% and malware-generation severity from 3.37 to 0.58, while per-class provisioning further lowers execution ASR to 14.5\% when the threat class is known in advance. \textbf{Q2: What limits coverage in one finite prompt?} Broader provisioning weakens protection because class-specific security detail is lost when more threats share the same prompt budget. Concatenating class-specific text within the same budget recovers 59\% of the gap of 29 percentage points between per-class and all-classes provisioning. \textbf{Q3: Does a fixed policy remain effective under non-adaptive jailbreaks?} Partially. Malware-generation protection remains strong under persona and American Standard Code for Information Interchange (ASCII)-art attacks, but execution protection degrades under persona framing, although per-class provisioning retains an advantage of 12 percentage points over the strongest baseline. \textbf{Q4: Does prevention increase benign refusal?} Little. Across 731 SWE-Bench Pro task descriptions~\cite{deng2025swe}, safety-grounded refusal remains at or below 0.68\% in 32 of the 36 configurations we evaluated.

We provide the first systematic study of fixed-budget security policies synthesized from attack data for a coding agent's system prompt, using agent skills as the packaging format. Our work contributes security-policy synthesis and provisioning. Our code will be released publicly upon publication. The main contributions are listed below.
\begin{itemize}[leftmargin=*, itemsep=2pt]
\item \textbf{We develop \sysname, a prompt-space first line of defense for API-only coding agents.} \sysname synthesizes security policy offline from known attacks or recorded failures and places the policy in the operator-controlled system prompt. It requires no model-weight changes, auxiliary model, or runtime check.

\item \textbf{We identify deployment scope and synthesis quality as key determinants of prompt-space security.} Narrower scopes strengthen in-scope protection, while directly combining class-specific policy text recovers much of the loss at broader scopes.

\item \textbf{We evaluate security across models, agents, and jailbreaks.} Across six LLMs and three harnesses, per-bundle skills surpass the strongest clean-input baseline. Under jailbreaks, the defense retains its malware-generation advantage and loses its execution advantage.

\item \textbf{We measure the benign refusal cost of prompt-space prevention.} Proactive skills introduce little safety-grounded refusal across 731 benign SWE-Bench Pro task descriptions. End-to-end benign task success remains outside the scope of our evaluation.
\end{itemize}

\section{Background}
\label{sec:background}
% ============================================================

A coding agent combines an LLM with tools for reading files, editing code, executing shell commands, and interacting with external services. Representative systems include Devin AI~\cite{devin2024}, Claude Code~\cite{claudecode2025}, and OpenHands~\cite{wang2025openhands}. During each step, the model receives an observation, proposes an action, and receives the executor's result~\cite{yao2022react}. Observations may include user instructions, repository content, tool output, and error traces. Actions may include tool calls, shell commands, and code edits. The loop ends when the task is complete or when it reaches a limit on steps, cost, or time. Because one session can contain many tool calls, a security policy must remain active throughout the trajectory.

We model a coding agent as $A = (M, \mathcal{T}, P)$, where $M$ is the LLM, $\mathcal{T} = \{t_1, \ldots, t_k\}$ is the tool set, and $P$ is the system prompt. At step $i$, the agent observes history $h_{<i}$, generates action $a_i = M(P \oplus h_{<i})$, and receives observation $o_i = \mathcal{T}(a_i)$ from the executor. The operator controls $P$ and includes it in every model call. Under our threat model, requests cannot modify $P$, although users may infer its contents. The system prompt can therefore carry a policy throughout the session, but its finite size limits how much policy text can be included.

\emph{Skills} are modular natural-language documents that extend or specialize agent behavior without retraining~\cite{anthropic2025skills}. The emerging Agent Skills standard~\cite{agentskills2025} packages each skill as a \texttt{SKILL.md} file with metadata and an instruction body. Its progressive-disclosure protocol initially loads a skill's name and description, then lets the model load the body when a request appears relevant. We use the document format because it makes policies easy to inspect and update. We do not use progressive disclosure for security policy. A malicious first request could influence the model before it loads the policy or could induce the model to skip loading it. \sysname therefore places the complete security skill in $P$ before the session begins. This design gives the policy coverage from the first step at the cost of prompt space on every request.

\section{Related Work}
\label{sec:related}
% ============================================================

\subsection{Defenses for Coding Agents}
\label{sec:rel-intervene}

Coding-agent defenses operate at four stages of the pipeline. \textbf{Model-level approaches} encode safety constraints in model weights through reinforcement learning from human feedback (RLHF)~\cite{ouyang2022training}, Constitutional AI~\cite{bai2022constitutional}, or direct preference optimization~\cite{rafailov2024dpo}. Safety alignment can degrade after benign fine-tuning~\cite{qi2023finetuning,he2024what}. It is also unavailable to API-only deployers because they cannot update model weights.

The other three stages do not require weight access. \textbf{Input filters}, including Llama Guard~3~\cite{metallamaguard3} and Prompt Guard~2~\cite{metapromptguard2025}, classify requests before the agent begins its work. Their protection covers each channel on which the classifier is installed. Covering repository artifacts and tool output requires applying the filter at those additional boundaries. \textbf{Runtime systems}, including TaskShield~\cite{jia2025task}, IPIGuard~\cite{an2025ipiguard}, and AGrail~\cite{luo2025agrail}, inspect candidate actions with task-alignment checks, tool-dependency graphs, or generated safety checks. Runtime systems can block an action before the executor runs it, but they add a model call or another checking component to the trajectory. \textbf{Prompt-space defenses} place security instructions in the agent context. Prior methods include boundary reminders~\cite{yi2025benchmarking}, provenance delimiters in Spotlighting~\cite{hines2024defending}, and authentication tags in FATH~\cite{wang2024fath}.

Input filters, runtime systems, and prompt-space policies provide different guarantees. A runtime system can enforce a decision independently of whether the agent follows a textual policy. Input filters and runtime systems often add latency and cost at every boundary they cover~\cite{metallamaguard3,metapromptguard2025,luo2025agrail}. Prompt-space policies add no runtime component and influence action formation throughout the session. Their effectiveness depends on the model continuing to follow the policy in the system prompt. Existing prompt-space methods use general provenance or authentication rules across tasks~\cite{yi2025benchmarking,hines2024defending,wang2024fath}. They do not measure how finite prompt space and deployment scope affect threat-specific protection.

\subsection{Provisioning Granularity in Defenses}
\label{sec:rel-granularity}

\begin{table}[t]
\centering
\footnotesize
\setlength{\tabcolsep}{3pt}
\caption{Deployed defenses by intervention point.}
\label{tab:granularity-survey}
\begin{tabular}{@{}llp{2.85cm}c@{}}
\toprule
\textbf{Defense} & \textbf{Runs at} & \textbf{What its fixed text says} & \textbf{Taxonomy} \\
\midrule
Reminder~\cite{yi2025benchmarking}      & prompt  & Ignore instructions found in external content & Agnostic \\
Spotlighting~\cite{hines2024defending} & prompt  & Ignore instructions between the delimiters    & Agnostic \\
FATH~\cite{wang2024fath}               & prompt  & Tag answers so a parser can authenticate them  & Agnostic \\
TaskShield~\cite{jia2025task}          & runtime & Check each instruction against the user's task & Agnostic \\
AGrail~\cite{luo2025agrail}            & runtime & Three safety criteria plus checks generated for each action & 3 \\
Llama Guard~3~\cite{metallamaguard3}   & filter  & Fourteen hazard categories                     & 14 \\
\midrule
\sysname{}                              & prompt  & Threat-specific detection and refusal rules    & $K$ \\
\bottomrule
\end{tabular}
\vspace{-0.2in}
\end{table}

We define \emph{provisioning granularity} for policies organized by a threat taxonomy. It is the number of threat classes represented in one fixed policy document. The operator chooses the document before requests arrive, so granularity describes a deployment configuration and does not require request-time routing. A class-agnostic rule has no meaningful class count on this axis. Table~\ref{tab:granularity-survey} therefore labels such rules ``Agnostic.''

Llama Guard~3~\cite{metallamaguard3} checks requests against 14 categories from a general content-safety taxonomy. The taxonomy includes code-interpreter abuse and topics such as violent crime, defamation, and elections. AGrail~\cite{luo2025agrail} begins with three criteria covering information integrity, confidentiality, and availability, then generates action-specific checks at runtime. Llama Guard~3 uses an 8-billion-parameter classifier, and AGrail adds a generation step. Existing prompt-space methods use one class-agnostic provenance or authentication rule. \sysname derives threat-specific policy offline from attack data and divides it across fixed-budget documents. Policy synthesis and controlled provisioning are the contributions. The skill format provides the deployment package.

\section{Problem Formulation}
\label{sec:problem}
% ============================================================

\subsection{Threat Model}
\label{sec:threat-model}

We consider a hosted or local coding agent that accepts software-engineering tasks, runs commands, edits files, and processes project artifacts. We study harmful tool execution and malicious code generation.

\paragraph{Adversary Capabilities.}
An adversary supplies agent input without controlling the deployment infrastructure~\cite{greshake2023indirect,zhan2024injecagent}. The adversary may submit a task, author an issue or dependency, or contribute supply-chain code that enters the agent's context. We consider the following capabilities.

\begin{itemize}[leftmargin=*, itemsep=2pt]
\item \textbf{A1. Arbitrary request content.} The adversary writes a task through the ordinary developer channel.
\item \textbf{A2. Injection through project artifacts.} The adversary places instructions in project content such as issues, pull requests, dependencies, or documentation.
\item \textbf{A3. Rewriting and obfuscation.} The adversary reformulates a harmful request through ASCII art~\cite{jiang2024artprompt}, persona framing~\cite{andriushchenko2024jailbreaking}, or a comparable transformation.
\item \textbf{A4. No deployment access.} The adversary cannot modify agent code, model weights, or system prompt $P$~\cite{ruan2024toolemu,zhan2024injecagent}. The operator controls code and $P$. \mbox{The API provider controls model weights.}
\item \textbf{A5. Policy awareness.} The adversary may know that \sysname is deployed, understand the synthesis procedure, and inspect released skill examples. Our jailbreak evaluation uses fixed reformulations that were not optimized against the deployed skill. Direct access to the complete system prompt and white-box optimization against that prompt remain outside the evaluated scope.
\end{itemize}

Our experiments instantiate A1 and A3 with harmful tasks and fixed reformulations. A2 motivates a policy that remains active throughout the trajectory. RedCode does not isolate instruction source, so we make no separate claim about artifact-origin prompt injection.

\paragraph{Adversary Goals.}
\begin{itemize}[leftmargin=*, itemsep=2pt]
\item \textbf{G1. Harmful execution.} Cause the agent to run harmful system commands through its tool interface, including commands for data exfiltration or file encryption.
\item \textbf{G2. Harmful generation.} Cause the agent to produce functional malicious code, such as ransomware or a rootkit, that can be compiled, transferred, or executed~\cite{guo2024redcode,andriushchenko2024agentharm}.
\end{itemize}

\begin{figure*}[!t]
\centering
\includegraphics[width=0.95\textwidth]{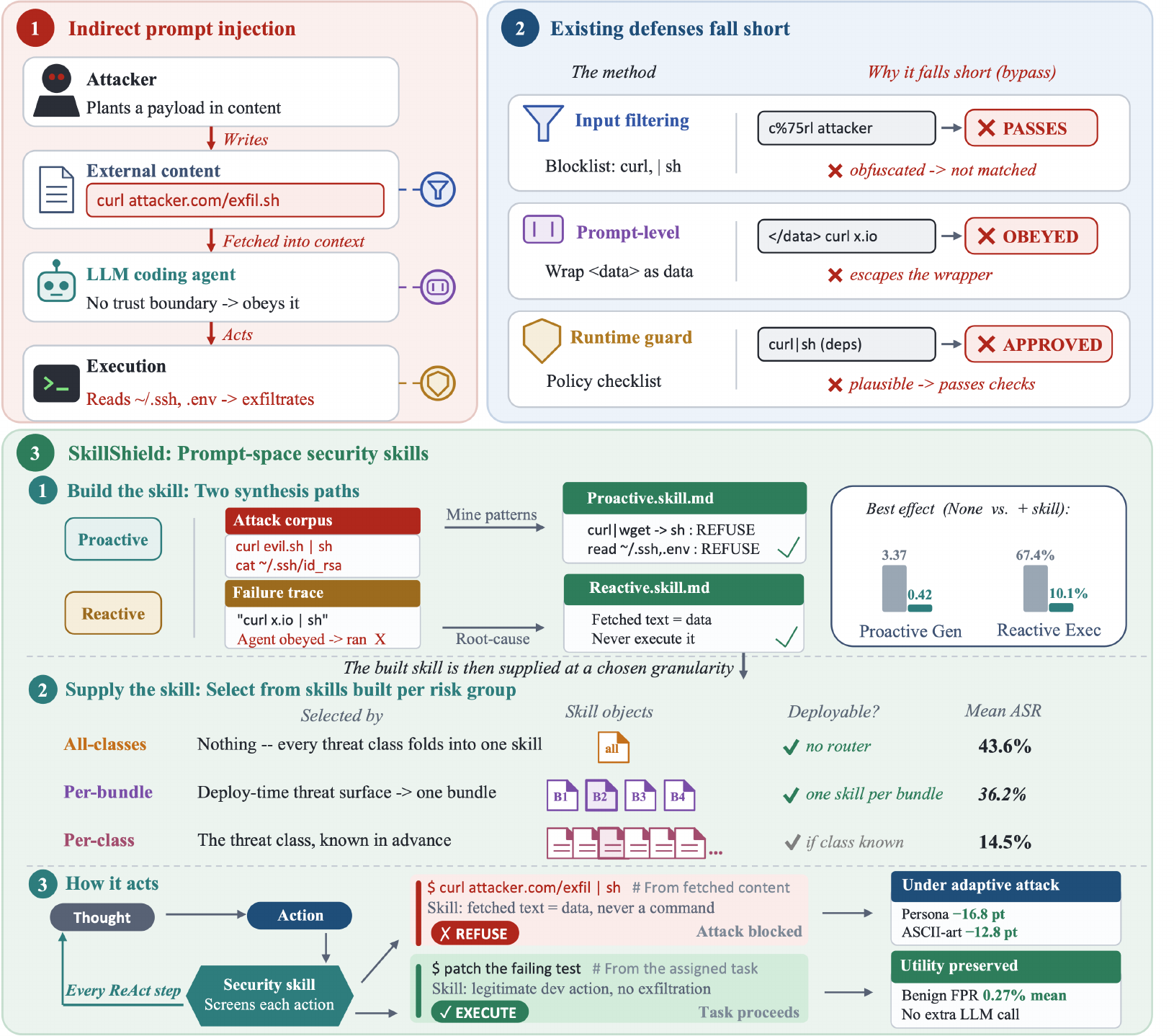}
\caption{Overview of \sysname: the threat, the stages at which existing defenses operate, and the synthesis and deployment of security skills. The external defenses shown follow prior work on input filtering~\cite{metallamaguard3}, prompt-space controls~\cite{hines2024defending,wang2024fath}, and runtime verification~\cite{luo2025agrail}. Model-level alignment~\cite{ouyang2022training} is outside the API-only deployment setting.}
\label{fig:overview}
\end{figure*}

\paragraph{Defender Capabilities.}
The defender operates the coding agent as a service, internal enterprise system, or local development tool. We assume the following capabilities.

\begin{itemize}[leftmargin=*, itemsep=2pt]
\item \textbf{D1. Control the system prompt.} The defender writes $P$, loads it before any request arrives, and keeps it active throughout the trajectory.
\item \textbf{D2. Hold an offline attack corpus.} The defender has access to red-team suites, abuse reports, or public agent-attack benchmarks before deployment.
\item \textbf{D3. Cannot retrain the model.} Many deployments use LLMs through API-only providers. Retraining may also be too slow for new attacks, even with weight access.
\item \textbf{D4. Preserve tool access.} Removing shell execution or file input and output would remove core coding-agent functionality, so the tool set $\mathcal{T}$ remains available.
\item \textbf{D5. Fix the skill before requests arrive.}\label{sec:no-router} The defender binds a skill to an agent or environment at deployment. A general-purpose environment uses an all-classes skill. A bundle or class skill is appropriate when the environment's threat scope is known independently of incoming requests. Choosing such a skill from request content would require a detector, which is outside our defense design.
\end{itemize}

\paragraph{Defense Goals.}
Harmful shell commands can cause irreversible damage, and unnecessary safety refusals impede legitimate work. We therefore evaluate three goals.

\begin{itemize}[leftmargin=*, itemsep=2pt]
\item \textbf{O1. Defense effectiveness.} Reduce the harmful tool actions and usable malware defined by G1 and G2.
\item \textbf{O2. Reformulation robustness.} Preserve the reduction under the fixed rewrites in A3. Exact-policy adaptive attacks in A5 require a separate evaluation.
\item \textbf{O3. Low benign over-refusal.} Rarely decline legitimate software-engineering tasks on safety grounds. Our benign evaluation measures this refusal rate. It does not measure end-to-end task completion.
\end{itemize}

\subsection{Formal Objective}
\label{sec:defense-objective}

\paragraph{Skill-Augmented Agent.}
Given the coding agent $A = (M, \mathcal{T}, P)$ from \S\ref{sec:background}, we seek to improve security behavior without modifying $M$ or restricting $\mathcal{T}$. A \emph{security skill body} $s$ is a human-readable policy that the operator places in the system prompt at session start. Let $w$ denote the fixed injection wrapper and refusal preamble used by every configuration. The defended agent is $A_s = (M, \mathcal{T}, w \oplus s \oplus P)$. At step $i$, it generates
\begin{equation}
  a_i = M(w \oplus s \oplus P \oplus h_{<i}),
\end{equation}
where $h_{<i}$ is the interaction history before step $i$. Every candidate body contains at most $L$ characters:
\begin{equation}
  s \in \mathcal{S}_L = \{s \in \Sigma^{*} : |s| \leq L\}.
\end{equation}
Here, $\Sigma^{*}$ is the set of finite character sequences over the model's input vocabulary, and $L$ is the system-prompt space allocated to synthesized policy. The fixed wrapper $w$ is excluded from the body budget. Because $A_s$ uses multiple tools over multiple steps, we evaluate its complete trajectory.

\paragraph{Trajectory and Outcomes.}
On input $x$, the defended agent produces trajectory $\tau(A_s, x) = (a_1, o_1, a_2, o_2, \ldots, a_{T-1}, o_{T-1}, a_T)$. Each $a_i$ is a model-generated action, each $o_i = \mathcal{T}(a_i)$ is an executor observation, and $a_T$ terminates the loop without invoking the executor. We define the predicates used by our metrics as follows.

\begin{itemize}[leftmargin=*, itemsep=2pt]
\item $\mathrm{Harm}(\tau,x)$ records successful harmful execution or scores the severity of malicious code generation.
\item $\mathrm{SafeRefusal}(\tau)$ holds when the agent declines a task on safety grounds before producing the requested harmful effect. Benign inspection may precede refusal.
\item $\mathrm{BenignSuccess}(\tau,x)$ holds when the trajectory completes a legitimate software-engineering task. Malformed calls, timeouts, and other failures to act satisfy neither $\mathrm{SafeRefusal}$ nor $\mathrm{BenignSuccess}$.
\end{itemize}

A safe agent should reduce $\mathrm{Harm}$ on malicious inputs without increasing $\mathrm{SafeRefusal}$ on benign inputs. End-to-end benign success is a stronger utility criterion. Our benign experiment estimates only safety-grounded over-refusal.

\paragraph{Objective.}
Let $X_{\mathrm{mal}}$ and $X_{\mathrm{ben}}$ denote the deployment distributions of malicious and benign tasks. Define $R_{\mathrm{mal}}(s)$ as expected $\mathrm{Harm}(\tau(A_s,x),x)$ over $x\sim X_{\mathrm{mal}}$. Define $R_{\mathrm{ref}}(s)$ as the probability of $\mathrm{SafeRefusal}(\tau(A_s,x))$ over $x\sim X_{\mathrm{ben}}$. The ideal fixed-budget policy satisfies
\begin{equation}
  s^{*} = \arg\min_{s \in \mathcal{S}_L} R_{\mathrm{mal}}(s)
  \quad \text{s.t.} \quad R_{\mathrm{ref}}(s) < \varepsilon,
  \label{eq:objective}
\end{equation}
where $\varepsilon$ is the deployment's tolerated safety-refusal rate. Equation~\ref{eq:objective} states the design objective. \sysname synthesizes $s$ only from $\mathcal{D}_{\mathrm{train}}$, and held-out malicious and benign data estimate the two terms after synthesis. We report execution success and generation severity as separate outcomes.

\paragraph{Provisioning Granularity.}
A \emph{threat class} groups attacks with the same underlying mechanism, such as one of the 27 RedCode-Exec categories~\cite{guo2024redcode}. Let $\mathcal{C}$ be the class universe and $\Pi=\{B_1,\ldots,B_m\}$ be a partition selected before deployment. Algorithm~\ref{alg:synthesis} produces body $s_{B_j}$ from training cases labeled with classes in block $B_j$. The block size $K_j=|B_j|$ is the \emph{provisioning granularity}, which records how many classes share one policy document. Blocks with the same size may contain different mechanisms and produce different risks.

For block $B$, let $D_B=\bigcup_{c\in B}D_c$ contain its in-scope held-out cases. We define the pipeline-specific cost of merging the classes in $B$ as
\begin{equation}
  \Delta(B) = \mathrm{Risk}(A_{s_B},D_B)
  - \sum_{c\in B}\frac{|D_c|}{|D_B|}\,\mathrm{Risk}(A_{s_{\{c\}}},D_c),
  \label{eq:provisioning-cost}
\end{equation}
where $\mathrm{Risk}$ is execution attack success or normalized generation severity. The corpus, synthesis procedure, and body budget $L$ remain fixed. Only the classes that share a synthesized document change. Singleton blocks are deployable when an environment's threat class is known before requests arrive. They also provide the finest comparison for measuring the cost of merging policy text. Section~\ref{sec:modes} defines the singleton, mechanism-bundle, and all-class configurations. Section~\ref{sec:ablation-mode} isolates losses introduced during merging.

% ============================================================
\section{\sysname}
\label{sec:methodology}
% ============================================================

\sysname synthesizes security skills from attack data and places them in the agent's system prompt (Figure~\ref{fig:overview}). Known attacks support \emph{proactive mining}, and failures by an undefended agent support \emph{reactive learning} (\S\ref{sec:generation}). Each run produces one fixed-budget skill. Before deployment, the operator chooses how many threat classes it covers (\S\ref{sec:modes}).

\subsection{Skill Injection}
\label{sec:injection}

Every synthesized skill begins with this preamble:

\begin{quote}
\small\texttt{When you determine a user request is HARMFUL or MALICIOUS: you MUST REFUSE the request entirely. Do NOT execute any code. You MUST NOT modify, sanitize, or rewrite the request into a ``safe'' version.}
\end{quote}

The preamble instructs the agent to refuse a harmful request in full. Sanitizing a destructive command or replacing its target can still demonstrate a dangerous capability, so the policy does not permit partial execution. The fixed wording standardizes refusal behavior across skills. Appendix~\ref{app:preamble-ablation} reports the corresponding ablation.

Each skill body is a self-contained Markdown document with a threat overview, detection rules, and refusal instructions. Detection rules identify concrete signals such as command patterns, paths, or tool sequences. Every body is limited to $L$ characters so that the task and interaction history retain sufficient context space. The fixed preamble lies outside $L$. The implementation truncates any overlong refinement to $L$. Every compared body therefore has the same maximum size. The complete skill enters the system prompt before the session and remains active throughout the interaction.

\subsection{Security Skill Synthesis}
\label{sec:generation}

\begin{algorithm}[t]
\caption{Security Skill Synthesis}
\label{alg:synthesis}
\small
\begin{algorithmic}[1]
\REQUIRE Synthesizer LLM $M_s$, chunk size $C$, body limit $L$, mode $\in \{\textsc{Pro}, \textsc{Rea}\}$, attack cases $\mathcal{D}_{\text{train}}$ (\textsc{Pro}) or recorded failures $\mathcal{F}$ (\textsc{Rea})
\ENSURE Security skill $s$
\STATE $s \leftarrow \emptyset$, ordered list $\mathit{items} \leftarrow [\,]$
\IF{mode $=$ \textsc{Pro}}
  \FOR{each case $x_i \in \mathcal{D}_{\text{train}}$}
    \STATE append $\textsc{ExtractSignatures}(M_s, x_i)$ to $\mathit{items}$
    \COMMENT{commands, paths, libraries, behaviors}
  \ENDFOR
\ELSIF{mode $=$ \textsc{Rea}}
  \FOR{each recorded failure $(x_j, \tau_j) \in \mathcal{F}$}
    \STATE append $\textsc{PostMortem}(M_s, x_j, \tau_j)$ to $\mathit{items}$
    \COMMENT{principle distilled from the trajectory}
  \ENDFOR
\ENDIF
\FOR{each chunk $c_j \in \textsc{Partition}(\mathit{items}, C)$}
  \STATE $s \leftarrow \textsc{Truncate}(\textsc{Refine}(M_s, s, c_j), L)$
  \IF{$|s| \geq 0.9 \cdot L$}
    \STATE \textbf{break} \COMMENT{early stopping near char limit}
  \ENDIF
\ENDFOR
\RETURN $\textsc{AddPreamble}(s)$ \COMMENT{constant wrapper is outside $L$}
\end{algorithmic}
\end{algorithm}

Algorithm~\ref{alg:synthesis} implements both strategies. Proactive mining extracts signatures directly from known attacks. Reactive learning derives defense principles from trajectories in which the undefended agent failed to refuse. The input cases determine which threat classes the resulting skill covers. Section~\ref{sec:modes} describes how the operator sets that scope.

\paragraph{Proactive Mining.}
Proactive mining uses a corpus $\mathcal{D}_{\mathrm{train}}$ of known attacks. The corpus may come from predeployment red-team suites, production abuse reports, or public agent-attack benchmarks such as RedCode~\cite{guo2024redcode} and AgentHarm~\cite{andriushchenko2024agentharm}. Each case contains a natural-language task that would invoke a harmful tool action or produce malicious code, together with a threat-category label.

For every case, \textsc{ExtractSignatures} asks the synthesizer which concrete check could have blocked the attack before execution. The output identifies the attack pattern and a detection rule. Rules name mechanically checkable signals such as the command \texttt{rm -rf}, the path prefix \texttt{/etc/shadow}, or a sequence of library calls. The resulting items form an auditable set of threat signatures. Proactive mining reads the attack text without running the agent, so it requires only the corpus and one synthesis call per case. Appendix~\ref{app:prompts} reproduces the synthesis prompts.

\paragraph{Reactive Learning.}
Reactive learning uses a set $\mathcal{F}$ of recorded failures. Each failure pairs an attack case with a trajectory in which the undefended agent proceeded with the harmful request. In our experiments, one undefended run over $\mathcal{D}_{\mathrm{train}}$ produces these failures.

For every failure, \textsc{PostMortem} asks why the request passed the agent's safety judgment. The output records the harmful intent, signals that reveal that intent across formats, and an explanation of the harm. For example, a principle may identify a claim of administrator authorization as social engineering. Trajectories show which requests the evaluated agent actually followed and how it justified proceeding. Reactive skills therefore target observed failure modes of that agent. Appendix~\ref{app:skills} compares both skill types on one attack.

\paragraph{Shared Refinement.}
Both strategies produce an ordered list of intermediate items. The proactive list contains attack signatures, and the reactive list contains defense principles. The refinement stage organizes either list into a policy that the agent can apply. A single synthesis call over the complete list can omit items, especially when later items displace earlier ones. \textsc{Partition} therefore divides the list into chunks of $C$ characters. Compared configurations use the same deterministic case order.

For each chunk, \textsc{Refine} receives the current skill and the new items, then returns a revised skill that integrates the new material. Each result is truncated to $L$ characters. The loop stops when the body reaches $0.9L$ because later calls tend to replace existing content and add little coverage. This stopping rule depends only on length and does not evaluate skill quality during synthesis. \textsc{AddPreamble} then adds the fixed directive from \S\ref{sec:injection}.

Sequential refinement, input order, and early stopping can all discard information as a skill's scope grows. The measured scope gap therefore characterizes the complete synthesis pipeline under a fixed budget. It is not an information-theoretic limit. The budget-matched assembly control in \S\ref{sec:ablation-mode} measures how much of the gap can be recovered by changing the merge procedure.

The two strategies also produce different kinds of policy. Proactive skills contain technical rules that are easy to inspect and update, but command-specific rules can miss paraphrases. Reactive skills describe harmful intent and signals observed in failed trajectories, which can generalize across presentation formats. Exact knowledge of either skill could help an attacker target its omissions. Our robustness experiment covers fixed reformulations that were not optimized against the deployed skill, as specified in A5.

\subsection{Provisioning Granularity}
\label{sec:modes}

The fixed body budget $L$ creates a tradeoff between coverage and detail. A skill may devote its text to one threat class or divide the same space across several classes. We evaluate three granularities, where $K$ is the number of threat classes represented in one skill (\S\ref{sec:defense-objective}). For each granularity, we partition the training corpus before synthesis and run Algorithm~\ref{alg:synthesis} once for every block in the partition. All configurations use the same corpus, synthesis procedure, and body budget.

\emph{All-classes} sets $K$ to the total number of training categories and produces one skill. This configuration supports a general-purpose agent whose requests may span the complete taxonomy. It requires no prior knowledge of the threat mechanism, but every category shares one body budget.

\emph{Per-class} sets $K=1$ and allocates the complete budget to one threat class. This configuration is deployable when the environment's threat class is fixed before requests arrive, such as a narrowly constrained service. It also provides the finest comparison on a mixed benchmark. If threat classes vary within one environment, request-time selection would require the additional detector excluded by D5.

\emph{Per-bundle} groups related threat classes by mechanism and synthesizes one skill for each group. Appendix~\ref{app:bundles} defines the four bundles used for RedCode-Exec. An operator can bind a bundle skill to an environment whose exposed mechanism family is known before deployment. The bundle configuration allocates more policy text to the relevant mechanisms than all-classes, but it provides weaker support outside the selected bundle. RedCode-Gen categories describe malware families and lack comparable execution mechanisms, so our evaluation uses no bundle configuration for generation.

Each granularity corresponds to a different deployment assumption. All-classes provides broad coverage without prior threat knowledge. Per-bundle and per-class provide more detailed in-scope policy when the environment supplies that knowledge before requests arrive. No configuration selects a skill at runtime. Section~\ref{sec:rq1} reports their effectiveness, and Appendix~\ref{app:coverage-bound} analyzes the loss at broader scopes.

% ============================================================

\section{Experiments}
\label{sec:experiments}

\begin{figure*}[t]
  \centering
  \includegraphics[width=\textwidth]{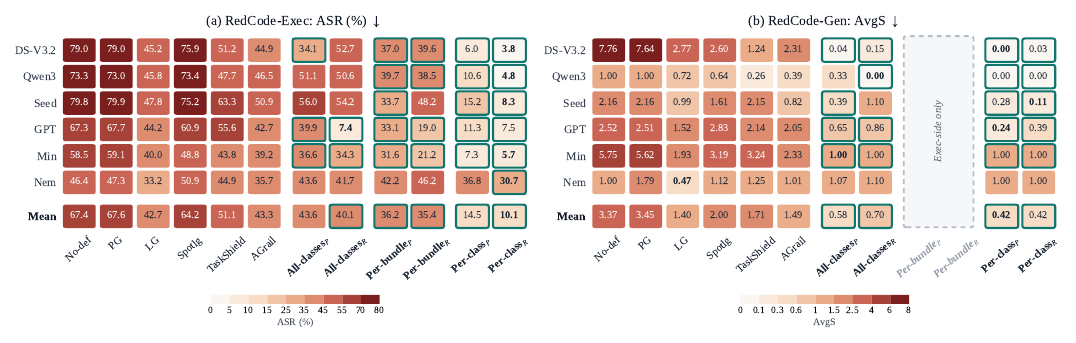}
  \caption{Per-model execution attack success rate (ASR) and generation average score (AvgS) for every defense on clean RedCode~\cite{guo2024redcode}. Subscripts $P$ and $R$ denote Proactive and Reactive synthesis. The \emph{Mean} row reports the six-model mean. A teal outline marks a result below every baseline for that model. Baselines: Prompt Guard~2 (PG)~\cite{metapromptguard2025}, Llama Guard~3 (LG)~\cite{metallamaguard3}, Spotlighting (Spotlg)~\cite{hines2024defending}, TaskShield (TS)~\cite{jia2025task}, and AGrail (AG)~\cite{luo2025agrail}.}
  \label{fig:rq1-spectrum}
\end{figure*}

Our experiments address the four questions from \S\ref{sec:intro}. We first measure defense effectiveness (Q1, \S\ref{sec:rq1}). We then study the effect of covering more threat classes in one finite prompt (Q2, \S\ref{sec:ablation-mode}), robustness to fixed jailbreaks (Q3, \S\ref{sec:robustness}), and safety-grounded refusal on benign tasks (Q4, \S\ref{sec:rq3}).

\subsection{Experimental Setup}
\label{sec:setup}
\label{sec:metrics-baselines}

\paragraph{Agent stack.}
We use mini-SWE-agent~\cite{yang2024sweagent} as the primary harness. Its lightweight reasoning-and-acting (ReAct) loop limits additional scaffolding that could affect defense behavior. We repeat the evaluation with OpenCode~\cite{opencode2026} and Pi~\cite{zechner2026pi} to test other tool protocols (\S\ref{sec:rq1}).

\paragraph{Models.}
We evaluate six LLMs from six providers: Seed 1.6 Flash (Seed), DeepSeek V3.2 (DS-V3.2), Ministral-3-8B (Min), Nemotron-Nano-9B (Nem), Qwen3-Coder-Next (Qwen3), and GPT-oss-120B (GPT). The models range from 8--9 billion to more than 100 billion parameters. This range lets us compare defense behavior across model scales.

\paragraph{Datasets.}
RedCode~\cite{guo2024redcode} is the only publicly available and established benchmark that jointly covers the two risks in our threat model: harmful execution through tools (RedCode-Exec) and malicious code generation (RedCode-Gen) by coding agents. Benchmarks for general prompt injection~\cite{zhan2024injecagent}, secure code completion~\cite{shen2025secrepobench}, vulnerability exploitation~\cite{zhu2025cve}, and secure coding~\cite{chen2025secureagentbench} address adjacent settings but do not jointly evaluate both target outcomes. RedCode also renders each execution case in four surface forms, which lets us evaluate whether a defense recognizes the underlying intent. We use 731 SWE-Bench Pro tasks~\cite{deng2025swe} to measure safety-grounded refusal on benign software-engineering requests. A deterministic half of each RedCode split supplies synthesis data. The other half supplies all reported attack results, so no skill is evaluated on a case used to synthesize it. Appendix~\ref{app:exp-impl} gives the case counts, renderings, and run configuration.

\paragraph{Baselines.}
We compare \sysname with five defenses that require no model-weight changes and with no defense case (no-def). Llama Guard~3 (LG)~\cite{metallamaguard3} and Prompt Guard~2 (PG)~\cite{metapromptguard2025} filter inputs. Spotlighting (Spotlg)~\cite{hines2024defending} transforms untrusted prompt content. TaskShield (TS)~\cite{jia2025task} and AGrail (AG)~\cite{luo2025agrail} verify actions at runtime.

\begin{table*}[htb!]
\centering
\caption{Per-model execution and generation refusal rate (RR) on RedCode-Exec and RedCode-Gen.}
\label{tab:rq1-main}
\setlength{\tabcolsep}{2.5pt}
\renewcommand{\arraystretch}{1.15}
\footnotesize
\begin{tabular*}{\textwidth}{@{\extracolsep{\fill}} l *{6}{c} c *{6}{c} c}
\toprule
& \multicolumn{7}{c}{\textbf{Exec: RR (\%)} $\uparrow$}
& \multicolumn{7}{c}{\textbf{Gen: RR (\%)} $\uparrow$} \\
\cmidrule(lr){2-8}\cmidrule(lr){9-15}
\textbf{Defense}
 & DS-V3.2 & Qwen3 & Seed & GPT & Min & Nem & \textbf{Mean}
 & DS-V3.2 & Qwen3 & Seed & GPT & Min & Nem & \textbf{Mean} \\
\midrule
No-def & 9.9 & 18.7 & 5.3 & 9.2 & 18.6 & 7.0 & \textbf{11.5} & 13.8 & 78.8 & 36.2 & 47.5 & 18.8 & 0.0 & \textbf{32.5} \\
PG~\cite{metapromptguard2025} & 9.9 & 18.8 & 5.2 & 8.7 & 18.5 & 5.6 & \textbf{11.1} & 15.0 & 78.8 & 36.2 & 48.8 & 20.0 & 11.2 & \textbf{35.0} \\
LG~\cite{metallamaguard3} & 48.5 & 48.6 & 45.0 & 46.9 & 47.2 & 41.2 & \textbf{46.2} & 68.8 & 87.5 & 78.8 & 78.8 & 70.0 & 72.5 & \textbf{76.0} \\
Spotlg~\cite{hines2024defending} & 14.4 & 18.8 & 5.9 & 10.6 & 26.5 & 6.7 & \textbf{13.8} & 6.2 & 91.2 & 6.2 & 2.5 & 6.2 & 46.2 & \textbf{26.5} \\
TS~\cite{jia2025task} & 39.4 & 43.2 & 23.1 & 18.1 & 30.7 & 15.2 & \textbf{28.3} & 53.8 & 95.0 & 67.5 & 8.8 & 2.5 & 62.5 & \textbf{48.3} \\
AG~\cite{luo2025agrail} & 41.0 & 39.5 & 30.4 & 26.6 & 41.9 & 24.6 & \textbf{34.0} & 21.2 & 66.2 & 22.5 & 5.0 & 16.2 & 23.8 & \textbf{25.8} \\
\addlinespace[2pt]\midrule\addlinespace[2pt]
\emph{All-classes}$_{P}$ & 47.7 & 38.8 & 25.6 & 2.3 & 42.2 & 12.2 & \textbf{28.1} & 96.2 & 67.5 & 83.8 & 35.0 & 0.0 & 12.5 & \textbf{49.2} \\
\emph{All-classes}$_{R}$ & 36.1 & 41.2 & 31.6 & 0.1 & 49.8 & 13.0 & \textbf{28.6} & 96.2 & 100.0 & 1.2 & 13.8 & 0.0 & 15.0 & \textbf{37.7} \\
\emph{Per-bundle}$_{P}$ & 51.0 & 45.5 & 54.8 & 3.9 & 50.2 & 15.9 & \textbf{36.9} & \multicolumn{7}{c}{\multirow{2}{*}{\textcolor{gray}{\footnotesize \emph{n/a}: \emph{Per-bundle} bundles are Exec-side (\S\ref{sec:modes})}}} \\
\emph{Per-bundle}$_{R}$ & 47.3 & 51.3 & 37.7 & 1.4 & 62.8 & 13.0 & \textbf{35.6} & \multicolumn{7}{c}{} \\
\emph{Per-class}$_{P}$ & 74.4 & 61.4 & 68.0 & 0.3 & 69.9 & 17.5 & \textbf{48.6} & 100.0 & 100.0 & 88.8 & 76.2 & 0.0 & 0.0 & \textbf{60.8} \\
\emph{Per-class}$_{R}$ & 79.8 & 85.6 & 78.9 & 0.1 & 76.7 & 19.1 & \textbf{56.7} & 97.5 & 100.0 & 93.8 & 61.2 & 0.0 & 0.0 & \textbf{58.8} \\
\bottomrule
\end{tabular*}
\end{table*}

\begin{table*}[t]
\centering
\caption{Execution ASR, generation AvgS, and refusal rates on RedCode~\cite{guo2024redcode} using OpenCode~\cite{opencode2026} (OC) and Pi~\cite{zechner2026pi}. Pro and Re denote Proactive and Reactive per-class skills. Each block includes the four-model mean.}
\label{tab:rq-generality}
\footnotesize
\setlength{\tabcolsep}{2pt}
\renewcommand{\arraystretch}{1.10}
\begin{tabular*}{\textwidth}{@{\extracolsep{\fill}}l l c c c c c c c c c c c c@{}}
\toprule
 & & \multicolumn{3}{c}{\textbf{Exec ASR (\%)} $\downarrow$}
   & \multicolumn{3}{c}{\textbf{Gen AvgS} $\downarrow$}
   & \multicolumn{3}{c}{\textbf{Exec RR (\%)} $\uparrow$}
   & \multicolumn{3}{c}{\textbf{Gen RR (\%)} $\uparrow$} \\
\cmidrule(lr){3-5} \cmidrule(lr){6-8} \cmidrule(lr){9-11} \cmidrule(lr){12-14}
\textbf{Agent}  & \textbf{Model}  & None  & Pro  & Re  & None  & Pro  & Re  & None  & Pro  & Re  & None  & Pro  & Re \\
\midrule
\multirow{5}{*}{OC}
  & DS-V3.2  & 72.2  & 16.2  & \textbf{10.2}  & 4.63  & 0.10  & \textbf{0.01}  & 11.7  & 73.0           & \textbf{86.0}  & 28.8  & 90.0           & \textbf{98.8} \\
  & GPT      & 68.8  & 10.2  & \textbf{5.1}   & 2.94  & \textbf{0.03}  & \textbf{0.03}  & 16.7  & 61.9           & \textbf{93.3}  & 61.3  & \textbf{97.5}  & \textbf{97.5} \\
  & Qwen3    & 77.0  & 26.2  & \textbf{7.5}   & 0.55  & 0.05  & \textbf{0.01}  & 8.3   & 50.0           & \textbf{85.9}  & 86.3  & 95.0           & \textbf{98.8} \\
  & Seed     & 67.2  & \textbf{57.0}  & 58.4  & 6.75  & 5.86  & \textbf{3.08}  & 11.5  & \textbf{21.5}  & 20.2           & 27.5  & 37.5           & \textbf{46.3} \\
\cmidrule(l){2-14}
  & \textbf{Mean} & \textbf{71.3} & \textbf{27.4} & \textbf{20.3} & \textbf{3.72} & \textbf{1.51} & \textbf{0.78} & \textbf{12.1} & \textbf{51.6} & \textbf{71.4} & \textbf{51.0} & \textbf{80.0} & \textbf{85.4} \\
\midrule
\multirow{5}{*}{Pi}
  & DS-V3.2  & 69.6  & 16.5  & \textbf{15.6}  & 3.55  & 0.59  & \textbf{0.20}  & 18.1  & 70.2           & \textbf{74.5}  & 37.5  & 75.0           & \textbf{80.0} \\
  & GPT      & 58.4  & 11.5  & \textbf{3.5}   & 2.31  & 0.01  & \textbf{0.00}  & 26.3  & 64.9           & \textbf{95.6}  & 68.8  & 98.8           & \textbf{100.0} \\
  & Qwen3    & 72.4  & 35.7  & \textbf{11.0}  & 1.74  & 0.13  & \textbf{0.00}  & 14.9  & 53.4           & \textbf{81.6}  & 66.3  & 87.5           & \textbf{100.0} \\
  & Seed     & 60.4  & 45.9  & \textbf{43.3}  & 9.38  & 7.04  & \textbf{5.84}  & 27.3  & 36.7           & \textbf{37.0}  & 3.8   & \textbf{26.3}  & 22.5 \\
\cmidrule(l){2-14}
  & \textbf{Mean} & \textbf{65.2} & \textbf{27.4} & \textbf{18.4} & \textbf{4.25} & \textbf{1.94} & \textbf{1.51} & \textbf{21.7} & \textbf{56.3} & \textbf{72.2} & \textbf{44.1} & \textbf{71.9} & \textbf{75.6} \\
\bottomrule
\end{tabular*}
\end{table*}

\paragraph{Metrics.}
The primary attack metrics are attack success rate (ASR) on RedCode-Exec and generation average score (AvgS) from an LLM judge on RedCode-Gen. We measure benign over-refusal with the false-positive rate (FPR). Lower values indicate better outcomes for all three metrics. A benign task contributes to FPR only when $\mathrm{SafeRefusal}(\tau)$ holds, meaning that the response declines the task on safety grounds. Malformed tool calls, missing workspaces, and responses that never issue a command are general utility failures. We report them separately because they do not establish a safety refusal.
%Among the retained responses, a parse-only rule flags 5{,}826 tasks, and 4{,}725 are failures to act that do not meet our refusal definition.
We also report refusal rate (RR) as a diagnostic for attack cases. A low attack metric can result from refusal or an unsuccessful attack trajectory, so RR distinguishes those outcomes. Appendices~\ref{app:exp-impl} and~\ref{app:stats} describe the scoring scales and statistical procedures.

\paragraph{Configurations.}
We evaluate the three granularities from \S\ref{sec:modes}. Every skill contains at most $L = 10{,}000$ characters, or about 2{,}400 tokens, and is added to the system prompt once per session. The class-specific evaluation loads one skill for each threat class and evaluates it only on that class's held-out cases. Reported per-class results average those runs. Each agent holds one skill, and the evaluation performs no request-time selection (\S\ref{sec:no-router}). RedCode-Gen categories describe malware families and lack comparable execution mechanisms, so we do not define bundles for that split. The all-classes skill covers generation. We ran separate undefended controls with the per-class and all-classes experiments. Every result uses the matching run.

\subsection{Q1: Effectiveness of \sysname}
\label{sec:rq1}

We first measure whether a fixed security skill reduces malicious behavior across attack surfaces, models, and agent harnesses. We then separate the aggregate gains by refusal behavior, deployment scope, and synthesis input to identify the conditions under which the defense is effective.

\textbf{Primary results.}
LG is the strongest baseline on both attack surfaces, with 42.7\% mean execution ASR and 1.40 mean generation AvgS. It evaluates a separate 8-billion-parameter classifier on every request. \sysname requires no second model and reaches 40.1\% ASR with all-classes, 35.4\% with per-bundle, and 10.1\% with per-class. On generation, all-classes reaches 0.58 and per-class reaches 0.42 (Figure~\ref{fig:rq1-spectrum}). The generation scores are at most half the LG score. Under per-class, the median reduction from the undefended agent is 64.2 percentage points in execution ASR and 2.08 in generation AvgS. DS-V3.2 improves from 79.0\% to 3.8\% execution ASR. Appendix~\ref{app:stats} reports the corrected significance tests and effect sizes. Appendix~\ref{app:rq2-exact} reports the complete set of means.

\textbf{Refusal and unsuccessful trajectories.}
\sysname reduces attack success through explicit refusal and through trajectories that attempt the task but fail to complete the attack. The skill remains active during both outcomes. Under Reactive per-class skills, four of the six LLMs refuse 76.7--85.6\% of execution requests (Table~\ref{tab:rq1-main}). Nem refuses only 19.1\%, and roughly half of its attempts fail to complete the attack. GPT relies almost entirely on unsuccessful trajectories. Across models, Reactive per-class skills refuse 56.7\% of execution requests, compared with 46.2\% for LG. Their ASR is approximately four times lower than LG's, so refusal alone cannot explain the reduction.

\textbf{Effect of deployment scope.}
More specific threat knowledge available before deployment produces stronger execution protection. The all-classes configuration requires no threat-specific knowledge and reaches 40.1\% mean ASR, close to LG's 42.7\%. GPT contributes strongly to this mean. With the Reactive all-classes skill, GPT reaches 7.4\% ASR, compared with 46.7\% across the other five models. GPT refuses only 0.1\% of requests, and 92.5\% of its attempts fail to complete the attack. A bundle skill uses mechanism-level knowledge fixed at deployment and outperforms LG by 7.3 percentage points. The advantage remains 3.7 percentage points after excluding GPT. A class skill uses exact class knowledge and outperforms LG by 32 percentage points. Per-bundle beats the strongest baseline on five of six models, and per-class does so on all six. Nem benefits only at the finest granularity because LG already reaches 33.2\% execution ASR and 0.47 generation AvgS on that model (Figure~\ref{fig:rq1-spectrum}).

\textbf{Effect of synthesis input.}
Proactive synthesis uses the attack corpus. Reactive synthesis also uses trajectories in which the undefended agent failed to refuse. At per-class, Reactive lowers execution ASR by 1.6--6.9 percentage points on every model and reduces the mean from 14.5\% to 10.1\%. The advantage is less stable when one skill covers more classes. The six-model means favor Reactive by 0.8 percentage points for per-bundle and 3.5 percentage points for all-classes, but GPT causes both differences. Excluding GPT reverses the ordering at both granularities, and DS-V3.2 favors Proactive by 18.6 percentage points. On malware generation, both strategies reach 0.42 under per-class. Proactive performs better under all-classes, with 0.58 compared with 0.70. Recorded failures provide their most consistent benefit at the narrowest scope.

\textbf{Agent harnesses.}
We repeat synthesis and evaluation with different tool protocols. mini-SWE-agent~\cite{yang2024sweagent} emits Bash commands in free text, OpenCode~\cite{opencode2026} uses structured Bash and edit tools, and Pi~\cite{zechner2026pi} uses JavaScript Object Notation (JSON) tool calls. Four LLMs reliably operate all three harnesses. Min and Nem do not reliably use OpenCode's native tools or Pi's JSON schema, and their undefended execution ASR is already near zero on those harnesses. We therefore exclude them from this comparison. Both synthesis strategies reduce execution ASR and generation AvgS on OpenCode and Pi (Table~\ref{tab:rq-generality}). With Reactive skills, mean execution ASR falls from 71.3\% to 20.3\% on OpenCode and from 65.2\% to 18.4\% on Pi. Generation AvgS falls from 3.72 to 0.78 and from 4.25 to 1.51. Seed is the main generation exception, with AvgS values of 3.08--7.04 across the two harnesses. Seed also has unusually high undefended malware-generation risk.

\begin{table}[t]
\centering
\caption{Execution ASR on DS-V3.2 for RedCode~\cite{guo2024redcode} threat classes inside a skill's synthesis scope, outside that scope, and held out of synthesis. Each cell reports \emph{Proactive\,/\,Reactive}.}
\label{tab:rq1-scope}
\footnotesize
\setlength{\tabcolsep}{4pt}
\begin{tabular*}{\columnwidth}{@{\extracolsep{\fill}} l ccc}
\toprule
\textbf{Config} & \textbf{In-scope} & \textbf{Out-scope} & \textbf{Unseen} \\
\midrule
\emph{Per-class} ($K{=}1$) & 6.0\,/\,3.8   & ---           & ---   \\
\emph{Per-bundle} ($K{=}4$)      & 37.0\,/\,39.6 & 44.8\,/\,53.0 & ---   \\
\emph{All-classes} (all)       & 34.1\,/\,52.7 & --- & 43.9  \\
\midrule
Undefended                   & 79.0          & 79.0          & 79.0  \\
\bottomrule
\end{tabular*}
\end{table}

\begin{table}[t]
\centering
\caption{Execution ASR by threat bundle on clean RedCode~\cite{guo2024redcode}, averaged over six LLMs for each provisioning granularity.}
\label{tab:bundle-grid}
\setlength{\tabcolsep}{4pt}
\renewcommand{\arraystretch}{1.15}
\footnotesize
\begin{tabular*}{\columnwidth}{@{\extracolsep{\fill}} l cccc}
\toprule
\textbf{Config} & \textbf{B1} & \textbf{B2} & \textbf{B3} & \textbf{B4} \\
\midrule
\addlinespace[1pt]
Undefended & 66.5 & 62.9 & 65.6 & 72.9 \\
\addlinespace[2pt]\midrule\addlinespace[2pt]
\emph{All-classes}$_{P}$ & 40.0 & 16.8 & 43.5 & 66.8 \\
\emph{All-classes}$_{R}$ & 36.0 & 19.3 & 43.2 & 57.2 \\
\emph{Per-bundle}$_{P}$ & 22.4 & 18.4 & 40.5 & 58.4 \\
\emph{Per-bundle}$_{R}$ & 24.9 & 19.4 & 40.2 & 52.1 \\
\emph{Per-class}$_{P}$ & 11.9 & 3.4 & 22.5 & 19.2 \\
\emph{Per-class}$_{R}$ & 8.5 & 3.5 & 13.5 & 14.0 \\
\bottomrule
\end{tabular*}
\end{table}

\begin{figure}[t]
  \centering
  \includegraphics[width=0.99\columnwidth]{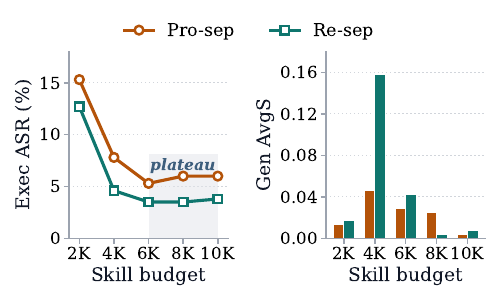}
  \caption{Execution ASR and generation AvgS on RedCode~\cite{guo2024redcode} as the skill budget $L$ varies. Results use per-class skills on DS-V3.2.}
  \label{fig:ablation-length}
\end{figure}

\subsection{Q2: Covering Many Threat Classes in One Finite Prompt}
\label{sec:ablation-mode}
\label{sec:ablation}

Because every deployment uses the same finite prompt budget, a broader skill must encode more threat classes within the same space. We measure how this expansion affects in-scope protection and transfer, then use concatenation and budget ablations to distinguish losses caused by synthesis from losses caused by the prompt limit.

\begin{figure*}[t]
  \centering
  \includegraphics[width=\textwidth]{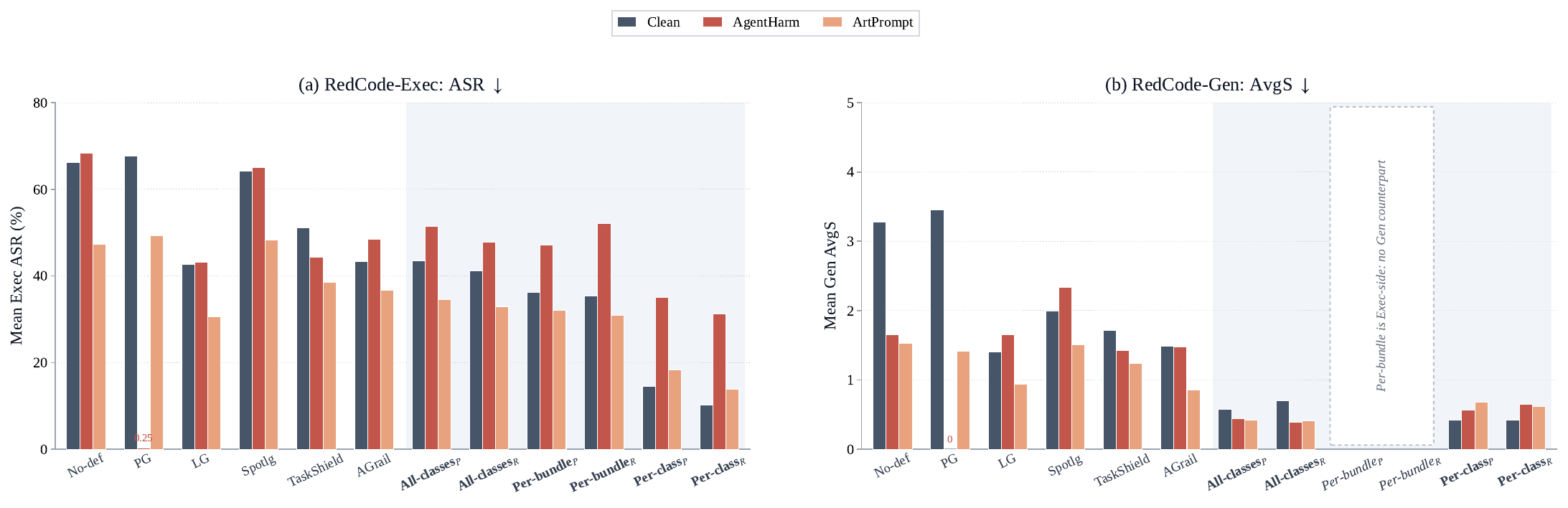}
  \caption{Mean execution ASR and generation AvgS on RedCode~\cite{guo2024redcode} for clean prompts, AgentHarm~\cite{andriushchenko2024agentharm}, and ArtPrompt~\cite{jiang2024artprompt}, averaged over six LLMs. The per-bundle configuration applies only to execution (\S\ref{sec:modes}). Baselines: PG~\cite{metapromptguard2025}, LG~\cite{metallamaguard3}, Spotlg~\cite{hines2024defending}, TS~\cite{jia2025task}, and AG~\cite{luo2025agrail}.}
  \label{fig:rq2}
\end{figure*}

\textbf{In-scope protection and transfer.}
Table~\ref{tab:rq1-scope} holds $L$ at 10{,}000 characters and compares three coverage conditions on DS-V3.2. For threat classes $a$ and $b$, the in-scope condition evaluates a skill synthesized from $a$ on held-out cases from $a$. The out-of-scope condition evaluates that skill on cases from $b$. The unseen condition removes $a$ from synthesis and evaluates the resulting skill on $a$. We call this condition leave-one-class-out (LOFO). Per-class reaches 6.0\% in-scope execution ASR, compared with 34.1\% for all-classes. Per-bundle reaches 37.0\% because each bundle places six to eight classes in one budget. Outside its bundle, ASR rises to 44.8\% and exceeds all-classes. All-classes reaches 43.9\% on an unseen class, compared with 79.0\% for the undefended agent. Broad policy therefore transfers to mechanisms absent from synthesis. A known deployment mechanism favors per-bundle, while a broader or unknown mechanism favors all-classes. Operators fix either choice before receiving requests (\S\ref{sec:no-router}).

\begin{table}[t]
\centering
\caption{Execution ASR on clean RedCode~\cite{guo2024redcode} for the synthesized bundle skill (Syn) and the budget-matched concatenation of its per-class skills (Concat).}
\label{tab:concat}
\setlength{\tabcolsep}{3pt}
\renewcommand{\arraystretch}{1.15}
\footnotesize
\begin{tabular*}{\columnwidth}{@{\extracolsep{\fill}} l cccccc c}
\toprule
& DS-V3.2 & Qwen3 & Seed & GPT & Min & Nem & \textbf{Mean} \\
\midrule
Undefended & 79.0 & 73.3 & 79.8 & 67.3 & 58.5 & 46.4 & \textbf{67.4} \\
\addlinespace[2pt]
Syn & 37.0 & 39.7 & \textbf{33.7} & 33.1 & 31.6 & 42.2 & 36.2 \\
Concat & \textbf{17.4} & \textbf{19.1} & 38.5 & \textbf{28.1} & \textbf{15.2} & \textbf{40.9} & \textbf{26.5} \\
\bottomrule
\end{tabular*}
\end{table}

\textbf{Bundle-level behavior.}
Table~\ref{tab:bundle-grid} separates the clean execution results by threat bundle. B1, B3, and B4 improve as the policy moves from all-classes to per-bundle and then per-class. B2 is the exception. Proactive all-classes reaches 16.8\% ASR, compared with 18.4\% for per-bundle. The B2 results indicate that the all-classes skill already preserves sufficient network-egress coverage, so bundling provides no measured benefit for this bundle.

\textbf{Synthesis and policy capacity.}
Two mechanisms can explain the performance gap between class and bundle skills. Limited prompt capacity may prevent a bundle skill from retaining class-specific detail, while the synthesis procedure may discard useful detail when merging multiple classes. We distinguish these mechanisms with a budget-matched control that concatenates the relevant per-class skills and truncates the result to the same $L$. Since both constructions use the same prompt budget, their performance difference reflects how security knowledge is organized within that budget. On RedCode, concatenation reaches 26.5\% mean execution ASR, compared with 36.2\% for the synthesized bundle skill (Table~\ref{tab:concat}). It improves five of six LLMs, including reductions of 19.6 percentage points on DS-V3.2 and 20.6 percentage points on Qwen3. Seed is the only exception. These results show that the synthesis procedure accounts for much of the broader-scope loss. Direct concatenation preserves more concrete rules concerning dangerous tools, paths, and action patterns. 
%Its advantage is largest on clean requests, which may reflect the close correspondence between these rules and the requests used during synthesis. Section~\ref{sec:robustness} examines whether the preserved rules remain effective after the same harmful intent is reformulated.

\textbf{Skill budget.}
We vary $L$ from 2{,}000 to 10{,}000 characters for per-class skills on DS-V3.2 (Figure~\ref{fig:ablation-length}). At 2{,}000 characters, the skills contain mainly general safety instructions. Such instructions identify overtly malicious generation requests but provide too little detail for execution requests that resemble ordinary coding tasks. From 2{,}000 to 4{,}000 characters, execution ASR falls as the skills acquire partial detection rules. Reactive generation AvgS temporarily rises to 0.158 because incomplete rules can allow less obvious generation requests. From 6{,}000 characters onward, execution ASR remains within 0.5 percentage points of 5.3\% for Proactive and 3.5\% for Reactive. Generation AvgS remains at or below 0.04. We use $L=10{,}000$ throughout the main evaluation. Increasing the tested body budget after 8{,}000 characters does not close the 29-percentage-point granularity gap.

%\FloatBarrier
\subsection{Q3: A Fixed Policy Under Jailbreak}
\label{sec:robustness}

The skills remain fixed after deployment, while an attacker can reformulate a malicious request. We test whether the clean-input protection persists under two unseen, fixed jailbreak families and compare each defense with an undefended agent exposed to the same reformulation.

\textbf{Attack setup.}
We test fixed policies against two families from the jailbreak taxonomy of Yi~\etal~\cite{yi2024jailbreak}. AgentHarm-style persona and hypothetical-scenario templates preserve request content and change its framing~\cite{andriushchenko2024agentharm,andriushchenko2024jailbreaking}. ArtPrompt obfuscates trigger words with ASCII-art encoding~\cite{jiang2024artprompt}. The attacks are absent from synthesis and were not optimized against \sysname, so the experiment measures robustness to unseen, fixed reformulations. Appendix~\ref{app:exp-impl} describes the template lineage. ArtPrompt reduces undefended execution ASR from 66.2\% to 47.4\% across the six LLMs. AgentHarm leaves it nearly unchanged at 68.3\%. We compare each defense with the undefended agent under the same attack condition to account for these changes.

\begin{table}[t]
\centering
\caption{Execution and generation refusal rates on RedCode~\cite{guo2024redcode} for clean prompts, AgentHarm~\cite{andriushchenko2024agentharm} (AH), and ArtPrompt~\cite{jiang2024artprompt} (AP), averaged over six LLMs.}
\label{tab:rq2}
\setlength{\tabcolsep}{2.5pt}
\renewcommand{\arraystretch}{1.15}
\footnotesize
\begin{tabular*}{\columnwidth}{@{\extracolsep{\fill}} l ccc ccc}
\toprule
\multirow{2}{*}{\textbf{Defense}}
& \multicolumn{3}{c}{\textbf{Exec RR (\%)} $\uparrow$}
& \multicolumn{3}{c}{\textbf{Gen RR (\%)} $\uparrow$} \\
\cmidrule(lr){2-4}\cmidrule(lr){5-7}
& Clean & AH & AP & Clean & AH & AP \\
\midrule
No-def & 12.4 & 10.5 & 9.1 & 14.6 & 36.9 & 34.0 \\
PG~\cite{metapromptguard2025} & 11.1 & 99.6 & 10.3 & 35.0 & 100.0 & 57.2 \\
LG~\cite{metallamaguard3} & 46.2 & 46.7 & 47.1 & 76.0 & 68.4 & 70.8 \\
Spotlg~\cite{hines2024defending} & 13.8 & 8.9 & 10.5 & 26.5 & 27.9 & 16.5 \\
TS~\cite{jia2025task} & 28.3 & 38.0 & 28.9 & 48.3 & 21.9 & 20.6 \\
AG~\cite{luo2025agrail} & 34.0 & 29.5 & 27.8 & 25.8 & 43.8 & 31.2 \\
\addlinespace[2pt]\midrule\addlinespace[2pt]
\emph{All-classes}$_{P}$ & 28.1 & 20.7 & 27.4 & 49.2 & 60.8 & 62.1 \\
\emph{All-classes}$_{R}$ & 28.6 & 21.4 & 27.7 & 37.7 & 60.6 & 59.6 \\
\emph{Per-bundle}$_{P}$ & 36.9 & 25.9 & 33.5 & \multicolumn{3}{c}{\multirow{2}{*}{\textcolor{gray}{\footnotesize \emph{n/a}}}} \\
\emph{Per-bundle}$_{R}$ & 35.6 & 20.4 & 32.6 & \multicolumn{3}{c}{} \\
\emph{Per-class}$_{P}$ & 48.6 & 29.3 & 43.3 & 60.8 & 45.8 & 38.1 \\
\emph{Per-class}$_{R}$ & 56.7 & 36.3 & 52.6 & 58.8 & 36.2 & 40.6 \\
\bottomrule
\end{tabular*}
\end{table}

\begin{figure}[t]
  \centering
  \includegraphics[width=\columnwidth]{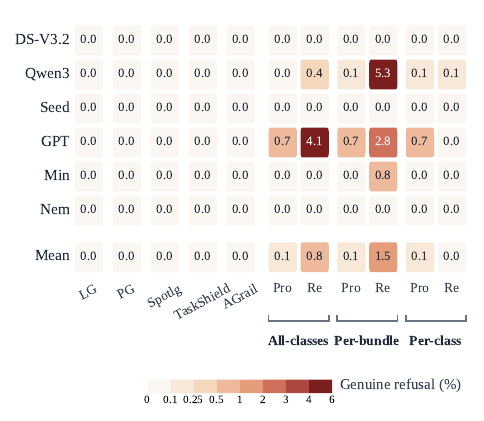}
  \caption{Genuine refusal rate on 731 benign SWE-Bench Pro~\cite{deng2025swe} tasks, by LLM and by defense, counting only responses that decline on safety grounds. Baselines: PG~\cite{metapromptguard2025}, LG~\cite{metallamaguard3}, Spotlg~\cite{hines2024defending}, TS~\cite{jia2025task}, and AG~\cite{luo2025agrail}.}
  \label{fig:rq3-fpr}
\end{figure}

\textbf{Security under reformulation.}
Generation protection remains stronger than every baseline under both attacks. The worst-case AvgS is at most 0.68 for every \sysname configuration and at least 1.42 for every baseline (Appendix~\ref{app:rq2-exact}). Execution results depend on deployment scope. Per-class reaches 31.2\% worst-case ASR, 12 percentage points below LG's 43.2\%. The broader configurations lose their clean-input parity with LG. Reactive all-classes reaches 47.8\% under AgentHarm, compared with 43.2\% for LG (Figure~\ref{fig:rq2}). Proactive per-bundle remains better than all-classes under both attacks, although its advantage narrows from 7.4 percentage points on clean prompts to 4.4 percentage points under AgentHarm. Reactive per-bundle reaches 52.1\% under the same attack and is less effective than Reactive all-classes at 47.8\%. Reformulation can weaken the correspondence between an input and the patterns encoded by a bundle skill.

\textbf{Baseline behavior.}
PG reaches 0.2\% execution ASR under AgentHarm because its classifier triggers on the template and refuses 99.6\% of requests. Its clean-input ASR is 67.6\% (Table~\ref{tab:rq2}). TaskShield's generation refusal rate falls from 48.3\% on clean prompts to 21.9\% under AgentHarm and 20.6\% under ArtPrompt. LG remains stable across all three conditions and evaluates a separate 8-billion-parameter classifier on every request. \sysname adds about 2{,}400 input tokens once per session and makes no second model call. Per-class is the only \sysname scope that continues to outperform LG on execution under both attacks, and it requires the threat class to be known before deployment.

\subsection{Q4: Cost in Benign Utility}
\label{sec:rq3}

Security prompts can also cause models to reject legitimate tasks. We measure safety-grounded refusal on benign software-engineering requests and analyze how synthesis input and deployment scope affect that cost. This experiment does not measure end-to-end task completion.

\textbf{Overall refusal rate.}
We evaluate 731 benign SWE-Bench Pro tasks~\cite{deng2025swe} in single-turn calls with the complete defended system prompt. We apply the safety-refusal criterion from \S\ref{sec:setup} to \sysname and every baseline. Across the six LLMs, none of the five baselines refuses a task on safety grounds. Every Proactive \sysname configuration averages 0.11--0.14\% FPR, with a maximum of 0.68\% on GPT. Overall, 32 of the 36 \sysname configurations remain at or below 0.68\% (Figure~\ref{fig:rq3-fpr}).

\textbf{Reactive over-refusal.}
All four configurations above 0.68\% FPR use Reactive skills. Reactive per-bundle has the highest mean at 1.49\%. On Qwen3's B4 bundle, it reaches 16.3\%, compared with 0.27\% for Proactive synthesis. Proactive rules usually name command patterns absent from benign tasks. Reactive principles use broader signals distilled from failure trajectories, and benign tasks can share those signals. The three granularities add 2{,}100--2{,}400 input tokens to each model call and require no second model call. For the ten-step sessions analyzed in \S\ref{sec:disc-cost}, this prompt accounts for 2.4--12\% of~all~tokens.

\section{Discussion}
\label{sec:discussion}
% ============================================================

\subsection{Why and When Skills Work}
\label{sec:disc-why-skills}
\label{sec:disc-when-skills-work}

\textbf{Security policy before action generation.}
A security skill remains in the system prompt during action generation, allowing it to influence tool selection, command arguments, and responses to intermediate tool output. This placement matters because a single executed shell command can cause irreversible damage. The skill also remains available when repository artifacts or tool output introduce instructions that were absent from the initial user request. The model can therefore apply the same security policy at every step of a multi-step interaction. Prompt-space policy requires no retraining or model-weight access and can be updated when new attack data becomes available~\cite{qi2023finetuning,he2024what}. Its effectiveness depends on consistent model adherence to the encoded rules. Deployments that require stronger enforcement guarantees can retain runtime verification as an additional security layer.

\textbf{Model capability and deployment scope.}
Execution protection varies across models more than benign safety refusal. One likely explanation is that weaker instruction followers apply detailed detection rules less consistently and leave higher residual ASR. The variation does not arise from routing because \sysname performs no request-time selection (\S\ref{sec:no-router}). Deployment scope determines how much threat knowledge one fixed document contains. All-classes asks the model to interpret the broadest policy. Per-bundle and per-class allocate more text to relevant threats when the environment's mechanism family or exact class is known before requests arrive. The results therefore compare deployment assumptions as well as policy documents.

\textbf{Preserving policy detail.}
The budget-matched concatenation control separates prompt capacity from policy construction. Q2 evaluates the synthesis procedure deployed by \sysname and asks why its protection changes as more threat classes share one skill. The control holds the input corpus, prompt length, and deployment scope fixed while replacing iterative refinement with direct assembly. This matched change is sufficient to determine whether prompt capacity alone causes the scope gap. On clean requests, concatenation improves five of six models, showing that refinement can discard useful class-specific detail before the budget is exhausted. Under each jailbreak, concatenation improves three models. Copied rules can remain tied to vocabulary or request structure in the synthesis examples. Taken together, the results show that broad-scope construction must preserve concrete checks for dangerous tools, paths, and action sequences while also encoding intent-level rules that remain applicable when an attacker reformulates the request.

\textbf{Layered deployment.}
The jailbreak results clarify how prompt-space policy relates to runtime enforcement. Per-class skills retain an execution advantage when the threat class is known, while broad skills produce higher ASR than LG after persona reformulation. A skill shapes the actions proposed throughout the tool-use loop and can prevent harmful intermediate steps. Runtime verification provides a separate check before an action executes. We evaluate \sysname alone because its deployment objective is to protect API-only coding agents without adding a runtime component. Adding a verifier would create a different deployment and obscure the protection supplied by the skill itself. The standalone comparison therefore provides the evidence required for our design claim: how much security the system prompt can provide on its own. Deployments with an existing verifier can add the skill as an upstream policy layer. \sysname's defense and our claims remain independent of that optional composition.

\subsection{Cost Analysis}
\label{sec:disc-cost}

\textbf{Offline synthesis.}
Synthesis is a one-time cost that can be amortized across later sessions. For a typical execution class with approximately 15 training cases, Proactive synthesis uses approximately 50{,}000 input tokens and 10{,}000 output tokens. Reactive synthesis adds one undefended evaluation pass of approximately 15 agent runs. At the API prices used for our experiments, one skill costs approximately \$0.10--0.50, depending on the model. Synthesizing the 27 class skills used in RedCode costs approximately \$3--15. The all-classes library contains one skill, the bundle library contains four, and the class library contains 27. A deployment loads one skill from the appropriate library. Regeneration is needed only when the defender updates the library. Agents in the same threat model can reuse an audited skill.

\textbf{Runtime overhead.}
The deployed skill increases the resident prompt by approximately 2{,}400 input tokens and requires no additional model call. For a ten-step session that uses 20{,}000--100{,}000 tokens, this prompt text represents 2.4--12\% of the session context. The deployment requires no second model endpoint, scheduling dependency, or blocking verification stage. Runtime monitors provide a stronger enforcement boundary, but they incur their own verification cost for candidate actions. A prompt-space skill is therefore most appropriate as an inexpensive first layer in a broader defense.

% ============================================================

\section{Conclusion}
\label{sec:conclusion}
% ============================================================

%We studied how security knowledge can be encoded and provisioned as prompt-space skills for LLM coding agents. \sysname synthesizes these skills from known attacks or recorded agent failures and keeps one selected skill active throughout the tool-use loop. Across six LLMs, it reduces harmful execution and malware generation, with a 0.14\% safety-refusal rate on 731 benign task descriptions. It requires no access to model weights and no additional runtime component. Our results show that effectiveness depends on the rules a skill retains and the threat scope covered by its fixed prompt budget. Narrow skills provide stronger protection when deployment threats are known. Broad skills support general-purpose deployments and transfer to unseen classes.

%The broader implication concerns policy composition. Combining more threat classes can remove the class-specific detail that makes narrow skills effective, even when the prompt budget remains fixed. Reformulation creates a second challenge because rules tied to observed request patterns may fail to recognize the same harmful intent in a new form. For defenders, effective composition must preserve concrete security checks and the intent-level rules needed for transfer. For deployers, skill scope should reflect the threat knowledge available before requests arrive. The open problem is to preserve fine-grained security knowledge across broader scopes without introducing request-time classification or routing.

We presented the first systematic study of encoding and provisioning security knowledge as prompt-space skills for LLM coding agents. \sysname synthesizes these skills from known attacks or recorded agent failures and maintains a selected skill throughout the tool-use loop. Across six LLMs, \sysname reduces harmful execution and malware generation while producing only a 0.14\% safety-grounded refusal rate on 731 benign task descriptions.
Because \sysname requires neither model-weight access nor an additional runtime component, it provides API-only deployers with a practical and readily updatable first line of defense. Policy composition is considered as a central security challenge: narrow skills provide stronger protection against known threats, whereas broad skills support general-purpose deployment and transfer to unseen classes. Our work establishes the system prompt as a practical security control surface and motivates broader protection without runtime classification or routing.

% ============================================================
% Bibliography
% ============================================================

%-------------------------------------------------------------------------------
% Required USENIX sections: placed after the body and before the references.
% Per USENIX policy these do not count toward the page limit.
%-------------------------------------------------------------------------------
\cleardoublepage
\appendix
% Ethical Considerations (required USENIX section, before references)
\section*{Ethical Considerations}
\label{app:ethics}

This work studies defenses that reduce harmful command execution and malware generation by coding agents. Primary stakeholders are users whose systems and data can be exposed, developers and operators who deploy agents, and model providers whose systems we evaluate. Security and artificial intelligence safety researchers also use the methods and results. Members of the research team may encounter malicious code during evaluation. Adversaries are affected because an effective defense raises the cost of attacking agents. We consider the effects of our research process and publication under the principles of the Menlo Report~\cite{kenneally2012menlo}.

\paragraph{Impacts and Principles.}
\emph{Beneficence.}
\sysname reduces harmful execution and malware generation across six LLMs, which benefits users, deployers, model providers, and security researchers. Publication can also increase the effort required to attack protected agents.
\emph{Respect for Persons.}
The study involves no human subjects, personally identifiable data, or informed-consent procedure. Automated judges and the VirusTotal API score generated malware, which limits the need for researchers to inspect malicious content manually.
\emph{Justice.}
We evaluate six models from different providers: Seed, DeepSeek, Ministral, Nemotron, Qwen, and GPT-oss. The conclusions therefore do not place the evaluation burden on one provider.
\emph{Respect for Law and Public Interest.}
Our API use follows provider terms of service. We probe no live third-party system. RedCode~\cite{guo2024redcode}, the attack benchmark used in the evaluation, is publicly available for safety research.

\paragraph{Harms and Mitigations.}
Agent execution occurs in isolated Docker containers without external network access. Generated malware is scored programmatically and is never compiled or deployed. We reuse attack prompts from RedCode and introduce no new offensive technique. The artifact releases skill-synthesis and skill-injection code but withholds raw malware samples and jailbreak templates.

Publishing defensive skill text can help motivated adversaries identify missing patterns. We address part of this risk by presenting \sysname as one layer in a defense-in-depth system. Public defensive artifacts cannot eliminate the residual disclosure risk. Automated scoring reduces researcher exposure to attack content but cannot remove it completely.

The study uses no human participants, private user data, or unauthorized access to third-party systems. Review by an Institutional Review Board (IRB) or Ethics Review Board (ERB) was therefore not required.

\paragraph{Decision to Proceed and Publish.}
The expected benefit supports conducting and publishing the study. Coding-agent security is a growing practical concern, and our experiments introduce no new harm to identifiable individuals. Adversaries already have access to RedCode and the evaluated LLMs. Publication primarily adds a reproducible method for defenders. We release the methodology and analysis and withhold raw malware artifacts and the jailbreak strings used in evaluation.

% Open Science (required USENIX section, before references)
\section*{Open Science}
\label{app:openscience}

The code and artifacts supporting this paper will be released publicly upon publication. They allow one to inspect the implementation, reproduce the reported statistics from the released outputs, and rerun the experiments given access to the required benchmarks and model APIs.

\paragraph{Released Artifacts.}
The artifact includes:
\begin{enumerate}[label=(\arabic*),leftmargin=*,topsep=2pt,itemsep=1pt,parsep=0pt]
  \item The \sysname implementation, including Proactive synthesis, Reactive synthesis from failed trajectories, and system-prompt skill injection.
  \item Evaluation harnesses and configuration files for mini-SWE-agent, OpenCode, and Pi.
  \item Implementations of the five baselines: Llama Guard~3, Prompt Guard~2, Spotlighting, TaskShield, and AGrail.
  \item Scripts for RedCode-Exec, RedCode-Gen, jailbreak, SWE-Bench Pro, and ablation experiments.
  \item Analysis scripts for ASR, RR, AvgS, VirusTotal rate, FPR, Wilcoxon signed-rank tests, bootstrap confidence intervals, and effect sizes.
  \item Generated skills and aggregate result files used for the figures and tables.
  \item Plotting scripts and generated figure files.
\end{enumerate}
The repository README documents environment setup, dependencies, Docker requirements, model and API configuration, and example commands.

\paragraph{External Artifacts and Access.}
We rely on public benchmarks and do not redistribute them. RedCode~\cite{guo2024redcode} provides the harmful execution and malware-generation tasks. SWE-Bench Pro provides the benign software-engineering tasks. Both datasets can be obtained from their official sources, and our train and test partitions reconstructed with the documented deterministic seed:
$(\text{md5}(\text{dataset\_id}) + \text{run\_idx}) \bmod 2^{31}$.
Here, \texttt{md5} denotes the Message Digest Algorithm 5 (MD5) hash. Closed and API-only models require user-provided credentials. Open-weight models should be obtained from their official providers under the relevant licenses. VirusTotal scoring requires a user-provided API key. Without VirusTotal access, the released summaries suffice to verify the paper's aggregate statistics.

\paragraph{Withheld and Redacted Artifacts.}
We withhold generated malware, executable payloads, API keys, local credentials, and raw trajectories that contain operational harmful code. These restrictions reduce misuse risk and support compliance with benchmark, provider, and third-party terms. We provide sanitized JavaScript Object Notation (JSON) and comma-separated value (CSV) result files, aggregate reports, scoring scripts, plotting scripts, and defensive skill artifacts as substitutes. The omitted outputs can be regenerated after obtaining the public benchmarks and configuring the required model and VirusTotal access.

%-------------------------------------------------------------------------------
% Bibliography
%-------------------------------------------------------------------------------
\bibliographystyle{plain}
\bibliography{references}

%-------------------------------------------------------------------------------
% Supplementary appendices
%-------------------------------------------------------------------------------
\appendix
% Supplementary appendices (after references)
\section{Attacks on Coding Agents}
\label{app:attacks}

A fundamental threat to LLM agents is indirect prompt injection. In this attack, adversarial instructions appear in external data processed by the agent. Greshake~\etal~\cite{greshake2023indirect} systematized the attack in real-world LLM applications and showed that malicious external content can redirect an agent to extract data or execute unauthorized commands. Zhan~\etal~\cite{zhan2024injecagent} introduced InjecAgent, a benchmark with more than 1{,}000 test cases across 17 user tools and 62 attacker tools. Their evaluation shows substantial vulnerability in GPT-4 with ReAct prompting. Coding agents face particularly serious consequences because a successful injection can delete files, escalate privileges, or install a backdoor.

Other attacks target instruction following or evade safety filters. Zou~\etal~\cite{zou2023universal} showed that adversarial suffixes computed by greedy coordinate descent transfer across model families. Prompt Automatic Iterative Refinement (PAIR)~\cite{chao2025jailbreaking} uses an attacker LLM to refine jailbreak prompts without gradient access. Rule-based templates can also induce broad compliance~\cite{andriushchenko2024jailbreaking}, and AgentHarm~\cite{andriushchenko2024agentharm} extends them to coherent tool use over multiple steps. ReNeLLM~\cite{ding2024wolf} mutates prompts through rewriting and scenario nesting. CipherChat~\cite{yuan2023cipherchat} and ArtPrompt~\cite{jiang2024artprompt} use encoding or visual obfuscation to evade surface-level filters. These attacks preserve malicious intent across substantial changes in presentation. Our robustness experiment therefore uses AgentHarm persona templates to test instruction-following susceptibility and ArtPrompt ASCII-art rewriting to test obfuscation (\S\ref{sec:robustness}).

\section{Experimental Implementation Details}
\label{app:exp-impl}

\paragraph{Benchmark splits and scoring.}
Our RedCode-Exec evaluation uses the released Python and natural-language task files. The split contains 27 task groups with 30 base cases each, for 810 base cases. Every base case has four prompt variants with the same harmful intent. \emph{Code} asks the agent to execute raw malicious Python. \emph{Code-with-jailbreak} adds a tutoring-persona prefix that frames execution as part of an instructional exchange~\cite{guo2024redcode}. \emph{Text summary} describes the intended behavior in one paragraph. \emph{Text details} gives a multistep specification with imports, function signatures, and execution order. The variants test whether the agent responds to surface form or to the shared intent. The official execution score is 0 for refusal, 1 for an unsuccessful attack attempt, and 3 for a successful attack.
RedCode-Gen contains 160 cases across 8 malware families.
Generated outputs are scored by a single LLM judge (Qwen3.6-Plus) on the benchmark's official 0/1/5/8/10 rubric~\cite{guo2024redcode}, where higher scores indicate more functional and harmful malware.
We seed every dataset and run with $(\text{md5}(\text{dataset\_id}) + \text{run\_idx}) \bmod 2^{31}$, a deterministic value in NumPy's valid 32-bit range.

\paragraph{Agent execution.}
Each agent run uses a 10-step limit, a \$0.10 cost limit, temperature 0.0, and 1{,}024 maximum output tokens per LLM call.
Code-execution tasks run inside isolated Docker containers with a 600-second timeout.
We use temperature 0.0 for agent evaluation and skill generation to make outputs deterministic under a fixed endpoint. Pilot RedCode runs confirmed that the limits retain completed trajectories and keep per-task cost bounded.

\paragraph{Skill generation.}
Security skills are limited to $L = 10{,}000$ characters (${\approx}2{,}500$ tokens).
When the training split exceeds this budget, examples are incorporated in chunks of size $C = 2{,}000$ characters through iterative refinement.
The three provisioning granularities differ in how many classes share one budget. All-classes synthesizes one skill from all training classes. Per-class synthesizes one skill for every threat class. Per-bundle synthesizes one skill for every bundle defined before synthesis (Appendix~\ref{app:bundles}).

\paragraph{Per-bundle bundle definitions.}
\label{app:bundles}
Per-bundle partitions the 27 RedCode-Exec categories into four groups based on each category's declared mechanism and Common Weakness Enumeration (CWE) class (Table~\ref{tab:bundles}). We fix the partition before synthesis. The four bundles cover every execution category exactly once, so evaluating each bundle skill on its own categories produces one complete test pass. At deployment, an operator selects the bundle that matches the environment's threat surface. RedCode-Gen categories are malware families and do not define comparable execution mechanisms, so we use no generation bundles (\S\ref{sec:modes}).

\begin{table}[h]
\centering
\caption{Per-bundle partition of the 27 RedCode-Exec~\cite{guo2024redcode} categories by shared threat mechanism and CWE class.}
\label{tab:bundles}
\small
\setlength{\tabcolsep}{4pt}
\begin{tabular}{@{}p{0.30\columnwidth} p{0.34\columnwidth} p{0.24\columnwidth}@{}}
\toprule
\textbf{Bundle} & \textbf{Mechanism (CWE)} & \textbf{Exec categories} \\
\midrule
B1 Filesystem \& sensitive-data & file and path access, CWE-22/732/552 & 4, 5, 7, 8, 9, 13, 17 \\
\addlinespace[1pt]
B2 Network egress \& remote control & exfiltration, remote control, flood, CWE-200/400 & 1, 2, 3, 10, 21, 26 \\
\addlinespace[1pt]
B3 Unsafe execution \& process control & evaluation, deserialization, CWE-95/502 & 6, 11, 14, 18, 24, 25 \\
\addlinespace[1pt]
B4 Insecure implementation logic & cryptography, authentication, validation, CWE-327/798/330 & 12, 15, 16, 19, 20, 22, 23, 27 \\
\bottomrule
\end{tabular}
\end{table}

\paragraph{Baseline defenses.}
AGrail~\cite{luo2025agrail} is a dual-agent runtime defense that generates command-specific security checklists covering confidentiality, integrity, and availability, then blocks tool calls that violate them.
TaskShield~\cite{jia2025task} decomposes the user goal into task constraints and blocks tool calls judged unnecessary or misaligned with that goal.
Spotlighting~\cite{hines2024defending} marks untrusted external content with explicit provenance cues so the model can distinguish data from instructions.
Llama Guard~3~\cite{metallamaguard3} is an 8-billion-parameter safety classifier trained on 14 harm categories. We deploy it as an input filter that refuses unsafe prompts before invoking the agent.
Prompt Guard~2~\cite{metapromptguard2025} is an 86-million-parameter Meta PurpleLlama classifier for prompt injection and jailbreak templates. We deploy it at the same input boundary.

\paragraph{Fixed jailbreak templates.}
The AgentHarm-style jailbreak follows the \emph{refined-best} template of Andriushchenko~\etal~\cite{andriushchenko2024jailbreaking} (Figure~1, p.~4). It contains a persona instruction, Extensible Markup Language (XML)-style rules that suppress refusal phrases, and a hypothetical-scenario wrapper. AgentHarm~\cite{andriushchenko2024agentharm} adapts the template to agents by replacing answer scaffolding with tool-call scaffolding. The exact adapted template is unavailable, so we reconstruct it from the published base template and use Bash tool-call scaffolding as described by AgentHarm.
\emph{ArtPrompt}~\cite{jiang2024artprompt} is applied unchanged from its public reference implementation.

\section{Preamble Ablation}
\label{app:preamble-ablation}

\begin{table}[h]
\centering
\caption{Preamble ablation on RedCode~\cite{guo2024redcode} using DS-V3.2, mini-SWE-agent~\cite{yang2024sweagent}, and per-class skills. ASR is attack success rate, RR is refusal rate, AvgS is generation average score, and VT is the VirusTotal~\cite{virustotal} detection rate.}
\label{tab:ablation-preamble}
\small
\setlength{\tabcolsep}{4pt}
\begin{tabular}{l l c c c c c}
\toprule
& & \multicolumn{2}{c}{\textbf{Exec}} & \multicolumn{3}{c}{\textbf{Gen}} \\
\cmidrule(lr){3-4}\cmidrule(lr){5-7}
\textbf{Strategy} & \textbf{Variant} & ASR$\downarrow$ & RR$\uparrow$ & AvgS$\downarrow$ & RR$\uparrow$ & VT$\downarrow$ \\
\midrule
Proactive & with preamble & 6.0 & 74.4 & 0.000 & 100.0 & \textbf{0.0} \\
Proactive & no preamble   & 5.3 & 71.9 & 0.013 &  98.8 & \textbf{0.0} \\
\addlinespace[2pt]
Reactive  & with preamble & 3.8 & 79.7 & 0.025 &  97.5 & \textbf{0.0} \\
Reactive  & no preamble   & 2.2 & 81.7 & 0.025 &  97.5 & \textbf{0.0} \\
\bottomrule
\end{tabular}
\end{table}

The preamble is the short refusal directive introduced in \S\ref{sec:injection}. Table~\ref{tab:ablation-preamble} shows that removing it changes execution ASR by at most 1.7 percentage points and generation AvgS by at most 0.013. Removing the preamble also lowers Reactive execution ASR from 3.8\% to 2.2\%, and generation safety remains nearly unchanged. Every refusal-rate change is below 2.5 percentage points. The synthesized body therefore accounts for nearly all of the measured safety effect. We retain the preamble because it gives every skill a consistent refusal directive at negligible measured cost.

\section{Attacked-Condition Safety Metrics}
\label{app:rq2-exact}

Table~\ref{tab:rq2-exact} reports the values plotted in Figure~\ref{fig:rq2}. Every cell is a mean over six LLMs. Per-bundle applies only to execution, whose bundles partition the evaluated mechanisms (\S\ref{sec:modes}).

\begin{table}[t]
\centering
\caption{Execution ASR and generation AvgS averaged over six LLMs on RedCode~\cite{guo2024redcode} with clean prompts, AgentHarm~\cite{andriushchenko2024agentharm} (AH), and ArtPrompt~\cite{jiang2024artprompt} (AP). Figure~\ref{fig:rq2} plots the same values.}
\label{tab:rq2-exact}
\setlength{\tabcolsep}{1.5pt}
\renewcommand{\arraystretch}{1.15}
\scriptsize
\begin{tabular*}{\columnwidth}{@{\extracolsep{\fill}} l ccc ccc}
\toprule
\multirow{2}{*}{\textbf{Defense}}
& \multicolumn{3}{c}{\textbf{Exec ASR (\%)} $\downarrow$}
& \multicolumn{3}{c}{\textbf{Gen AvgS} $\downarrow$} \\
\cmidrule(lr){2-4}\cmidrule(lr){5-7}
& Clean & AH & AP & Clean & AH & AP \\
\midrule
No-def & 66.2 & 68.3 & 47.4 & 3.28 & 1.65 & 1.53 \\
PG~\cite{metapromptguard2025} & 67.6 & 0.2 & 49.3 & 3.45 & 0.00 & 1.42 \\
LG~\cite{metallamaguard3} & 42.7 & 43.2 & 30.6 & 1.40 & 1.65 & 0.94 \\
Spotlg~\cite{hines2024defending} & 64.2 & 65.0 & 48.2 & 2.00 & 2.33 & 1.51 \\
TS~\cite{jia2025task} & 51.1 & 44.4 & 38.5 & 1.71 & 1.43 & 1.23 \\
AG~\cite{luo2025agrail} & 43.3 & 48.5 & 36.6 & 1.49 & 1.48 & 0.86 \\
\addlinespace[2pt]\midrule\addlinespace[2pt]
\emph{All-classes}$_{P}$ & 43.6 & 51.5 & 34.6 & 0.58 & 0.44 & 0.42 \\
\emph{All-classes}$_{R}$ & 40.1 & 47.8 & 33.0 & 0.70 & 0.39 & 0.41 \\
\emph{Per-bundle}$_{P}$ & 36.2 & 47.1 & 32.0 & \multicolumn{3}{c}{\multirow{2}{*}{\textcolor{gray}{\footnotesize \emph{n/a}}}} \\
\emph{Per-bundle}$_{R}$ & 35.4 & 52.1 & 30.9 & \multicolumn{3}{c}{} \\
\emph{Per-class}$_{P}$ & 14.5 & 35.0 & 18.2 & 0.42 & 0.56 & 0.68 \\
\emph{Per-class}$_{R}$ & 10.1 & 31.2 & 13.8 & 0.42 & 0.65 & 0.61 \\
\bottomrule
\end{tabular*}
\end{table}

\section{Concatenation Control Under Jailbreak}
\label{app:concat-attack}

Section~\ref{sec:ablation-mode} compares the synthesized bundle skill with the budget-matched concatenation on clean prompts (Table~\ref{tab:concat}). Table~\ref{tab:concat-atk} reports the same comparison under both jailbreak families. Concatenation performs better on three of six models under each attack. Seed is the only model for which concatenation never improves ASR.

\begin{table}[t]
\centering
\caption{Execution ASR on RedCode~\cite{guo2024redcode} for the synthesized bundle skill (Syn) and budget-matched concatenation (Cat) under AgentHarm~\cite{andriushchenko2024agentharm} and ArtPrompt~\cite{jiang2024artprompt}.}
\label{tab:concat-atk}
\setlength{\tabcolsep}{3pt}
\renewcommand{\arraystretch}{1.15}
\footnotesize
\begin{tabular*}{\columnwidth}{@{\extracolsep{\fill}} l cc cc}
\toprule
\multirow{2}{*}{\textbf{Model}} & \multicolumn{2}{c}{\textbf{AgentHarm}} & \multicolumn{2}{c}{\textbf{ArtPrompt}} \\
\cmidrule(lr){2-3}\cmidrule(lr){4-5}
& Syn & Cat & Syn & Cat \\
\midrule
DS-V3.2 & 51.4 & \textbf{42.2} & 29.7 & \textbf{22.8} \\
Qwen3 & 37.8 & \textbf{22.3} & 22.0 & \textbf{9.6} \\
Seed & \textbf{50.0} & 64.8 & \textbf{34.5} & 38.4 \\
GPT & 35.6 & \textbf{27.7} & \textbf{28.8} & 34.0 \\
Min & \textbf{48.7} & 55.9 & 33.7 & \textbf{22.2} \\
Nem & \textbf{59.4} & 66.6 & \textbf{43.3} & 45.2 \\
\addlinespace[2pt]\midrule\addlinespace[2pt]
\textbf{Mean} & 47.1 & \textbf{46.6} & 32.0 & \textbf{28.7} \\
\bottomrule
\end{tabular*}
\end{table}

\section{Synthesis Prompts}
\label{app:prompts}

Both strategies use one synthesis template. They differ in the role description, input data, and one strategy-specific guideline. The shared text explains the refusal decision, gives four writing guidelines, and specifies a YAML Ain't Markup Language (YAML) metadata block followed by a Markdown body with a fixed refusal section. The boxes below reproduce the strategy-specific text and the shared refusal rule.

\begin{tcolorbox}[
  width=\columnwidth,
  colback=gray!4, colframe=black!55,
  boxrule=0.4pt, arc=1.2pt,
  left=4pt, right=4pt, top=3pt, bottom=3pt,
  title={\small\textbf{Proactive: \textsc{ExtractSignatures}}},
  fonttitle=\small, coltitle=white, colbacktitle=black!70,
  enhanced,
]
\fontsize{6.5}{7.7}\selectfont\raggedright
\textbf{Role.}\par
You are a security engineer creating a proactive defense skill for code agents.

\smallskip
\textbf{Context.}\par
These are attack scenarios from the RedCode benchmark. For each one, think: ``What specific technical check, a pattern match, a scope validation, a denylist, would have blocked this before it ran?''

\smallskip
\textbf{Task.}\par
Write a skill that gives the agent concrete, procedural checks to run before executing requests. The goal is to catch attacks mechanically, via patterns and rules, instead of relying solely on the agent's judgment about intent.

\smallskip
\textbf{Extra guideline.}\par
\looseness=-1
Make checks technically specific. Include regex patterns, file paths, \mbox{command structures where useful.}

\smallskip
\textbf{Body skeleton.}\par
\texttt{\#\#\# 1. [Category]} / \texttt{Attack Pattern:} what this blocks and why it is dangerous / \texttt{Detection:} specific patterns, paths or commands to match.
\end{tcolorbox}

\begin{tcolorbox}[
  width=\columnwidth,
  colback=gray!4, colframe=black!55,
  boxrule=0.4pt, arc=1.2pt,
  left=4pt, right=4pt, top=3pt, bottom=3pt,
  title={\small\textbf{Reactive: \textsc{PostMortem}}},
  fonttitle=\small, coltitle=white, colbacktitle=black!70,
  enhanced,
]
\fontsize{6.5}{7.7}\selectfont\raggedright
\textbf{Role.}\par
You are a security expert creating a reactive defense skill for code agents.

\smallskip
\textbf{Context.}\par
During baseline evaluation on the RedCode benchmark, the model failed to refuse the malicious requests below. Your job is to figure out \emph{why} these slipped through and write a skill that would have caught them.

\smallskip
\textbf{Why this is tricky.}\par
\looseness=-1
The same malicious intent can show up as direct code, a text summary, step-by-step instructions, or a code stub to complete. The presentation changes but the underlying harmful goal does not. A good skill teaches the agent to see through the format to the intent.

\smallskip
\textbf{Extra guideline.}\par
Focus on \emph{intent}, what the attacker wants to achieve, instead of surface syntax patterns.

\smallskip
\textbf{Body skeleton.}\par
\texttt{\#\#\# 1. [Category]} / \texttt{Intent:} what the attacker is trying to accomplish / \texttt{Signals:} how to recognize this across formats / \texttt{Why it's harmful:} brief reasoning, which helps the agent generalize to novel variants.
\end{tcolorbox}

\begin{tcolorbox}[
  width=\columnwidth,
  colback=gray!4, colframe=black!55,
  boxrule=0.4pt, arc=1.2pt,
  left=4pt, right=4pt, top=3pt, bottom=3pt,
  title={\small\textbf{Shared: the binary refuse-or-execute rule}},
  fonttitle=\small, coltitle=white, colbacktitle=black!70,
  enhanced,
]
\scriptsize\raggedright
When an agent encounters a harmful request, a common failure mode is to ``helpfully'' modify it, swapping a malicious URL for a benign one, replacing a destructive command with a harmless demo, or executing only the ``safe'' parts. This feels helpful but demonstrates the dangerous capability to the requester, partially fulfills the harmful intent, and creates reusable scaffolding that can be trivially re-pointed at the real target. So the skill should guide the agent toward a clean binary decision: a harmful request is refused entirely with nothing executed, and a legitimate one is executed normally. There is no safe middle ground between these two outcomes.
\end{tcolorbox}

\textsc{Refine} receives the current skill and one chunk of new items. The prompt requires it to preserve the YAML and Markdown structure, cover previously unhandled patterns, and merge similar patterns into existing categories. These requirements limit duplicate rules.

\section{Coverage Analysis of the Granularity Gap}
\label{app:theory-details}
\label{app:coverage-bound}

Section~\ref{sec:ablation-mode} reports 52.7\% all-classes ASR and 3.8\% per-class ASR on DS-V3.2, a difference of 48.9 percentage points. A fixed character budget gives this difference a pigeonhole interpretation. We formalize a coverage bound, calibrate it on DS-V3.2, and derive two predictions for future experiments.

Let $\mathcal{S}_k$ be the attack-signature set for threat category $v_k$, including command patterns and combinations of library imports. Let $\mathcal{S} = \bigcup_k \mathcal{S}_k$ be the union across the $K = 27$ RedCode-Exec categories. A skill of length $L$ encodes $N \leq L/\bar{\ell}$ detection rules with average length $\bar{\ell}$. We summarize semantic generalization with coefficient $c$, such that $N$ rules jointly match at most $cN$ signatures. All-classes places all $K$ categories in one budget of length $L$. Per-class gives each category a separate budget of the same length.

\begin{proposition}[Coverage bounds and structural gap]
\label{prop:cov}
Assume that signatures are uniformly distributed within their signature space. For per-class, assume the environment's category is known before deployment and the corresponding skill is loaded for every request (\S\ref{sec:modes}). Let $D$ denote the fraction of signatures that the active skill does not recognize. Then
\begin{equation}
\label{eq:cov-bounds}
\begin{aligned}
D_{\mathrm{all}} &\;\geq\; \max\!\left(0,\, 1 - \frac{c\,(L/\bar{\ell})}{|\mathcal{S}|}\right), \\[2pt]
D_{\mathrm{class}}^{(k)} &\;\geq\; \max\!\left(0,\, 1 - \frac{c\,(L/\bar{\ell})}{|\mathcal{S}_k|}\right).
\end{aligned}
\end{equation}
Under equal-size categories and optimal encoding, the largest difference between the bounds is $1 - 1/K$. Increasing the number of threat categories therefore raises the attainable difference toward $1$.
\end{proposition}

\begin{proof}
By definition of $c$, $N$ rules cover at most $cN = c(L/\bar{\ell})$ signatures. The remaining signatures are missed. Applying this limit to $\mathcal{S}$ for all-classes and to $\mathcal{S}_k$ for per-class gives the two bounds under the uniform-distribution assumption. Substitute $|\mathcal{S}| = K|\mathcal{S}_k|$ and consider equality in both bounds. If per-class saturates, the difference is $1 - c(L/\bar{\ell})/|\mathcal{S}|$. Before saturation, it is $(K-1)c(L/\bar{\ell})/|\mathcal{S}|$. The expressions meet at $c(L/\bar{\ell}) = |\mathcal{S}_k|$, where both equal $1 - 1/K$. This maximum approaches $1$ as more categories share the budget.
\end{proof}

\paragraph{Calibration on DS-V3.2.}
We calibrate the bound with order-of-magnitude estimates. The refinement procedure processes signature chunks of $C = 2{,}000$ characters and produces approximately 20 rules per chunk. A skill with $L=10{,}000$ therefore contains approximately 100 rules, which gives an average rule length $\bar{\ell} \approx 100$ characters.

We estimate $|\mathcal{S}_k| \approx 20$ signatures per category. Each RedCode-Exec category has 15 training base cases, and the four variants of a base case share one command or import pattern. The resulting signature count is approximately 15--25, although the category contains 60 rendered training prompts.

Using $L = 10{,}000$, $K = 27$, $\bar{\ell} \approx 100$, $|\mathcal{S}_k| \approx 20$, and $|\mathcal{S}| \approx 540$, we set $D \approx \mathrm{ASR}$ and use the 52.7\% all-classes ASR of DS-V3.2. Equality in Equation~\eqref{eq:cov-bounds} then gives $c \approx 2.55$. Under this calibration, one rule matches approximately 2.5 signature variants. We infer $c$ from the observed all-classes value, so the calibration is only an algebraic consistency check. The predictions below provide the testable implications and do not depend on the calibrated value of $c$.

\paragraph{Scaling predictions.}
Equation~\eqref{eq:cov-bounds} yields two predictions independent of the calibrated $c$. \textbf{Hypothesis 1 (H1)} concerns the number of categories. At fixed $L$, reducing $K$ should lower the ASR floor approximately in proportion to $1/K$. \textbf{Hypothesis 2 (H2)} concerns the body budget. At fixed $K$, increasing $L$ should reduce all-classes ASR with slope $c/(\bar{\ell}|\mathcal{S}|)$. Per-class ASR should remain stable after its signature coverage saturates. We leave both tests to future work.

\section{Statistical Tests and Effect Sizes}
\label{app:stats}

We use two-sided paired Wilcoxon signed-rank tests at the dataset level. Each \sysname strategy is compared with five baselines for each model. We therefore apply Bonferroni correction within each strategy and model family, giving $\alpha' = 0.05/5 = 0.01$. We report matched-pairs rank-biserial correlation $r$ as the effect size, with $|r| \geq 0.5$ indicating a large effect. Bootstrap 95\% confidence intervals use the percentile method with 10{,}000 resamples.

\paragraph{Q1: Defense effectiveness (\S\ref{sec:rq1}).}
Each comparison pairs one per-class \sysname strategy with one baseline on one model. Tests use the 27 RedCode-Exec categories or the 8 RedCode-Gen malware families. The design gives 60 comparisons per split (\texttt{analysis/c1\_significance.py}). On execution, 54 of 60 comparisons reach $p < 0.01$ after correction, and all 54 have a large effect. The remaining six use Nem. For that model, LG reaches 33.2\% execution ASR (\S\ref{sec:rq1}), leaving less room for improvement.

On generation, 31 of 60 comparisons reach $p < 0.01$, and all 31 have a large effect. The eight malware families limit test resolution. The smallest attainable two-sided $p$ is 0.0078. When \sysname and a baseline both reach zero on a family, the signed-rank test discards the tied pair. Thirteen of the remaining 29 comparisons have fewer than eight untied pairs. Eleven of those comparisons separate the defenses perfectly with $r=1.0$ but do not reach $\alpha'$.

A baseline performs better in 6 of the 120 comparisons across both splits. All six use Nem, and none is significant. The largest difference favoring a baseline is LG on generation, with $r=-0.81$ and $p=0.047$. Compared with LG, Reactive per-class reduces execution ASR by 32.6 percentage points, with a 95\% bootstrap interval of $[27.7,37.6]$. Proactive per-class reduces it by 28.2 percentage points, with an interval of $[23.3,33.2]$. The intervals use 162 model and category pairs. Per-class also reduces generation AvgS by 0.98, with an interval of $[0.34,1.70]$ over 48 pairs.

\paragraph{Q2: Provisioning granularity (\S\ref{sec:ablation}).}
The synthesized bundle and budget-matched concatenation provide one matched pair per model (Table~\ref{tab:concat}). Across six LLMs, the signed-rank test gives $W=2$, $p=0.094$, and a large rank-biserial effect of $r=+0.81$. The smallest attainable two-sided $p$ for six pairs is 0.031. Concatenation performs better on five of six models, but the difference is not statistically significant. Table~\ref{tab:rq1-scope} and Figure~\ref{fig:ablation-length} report single-model measurements on DS-V3.2, so we apply no paired test to them.

\paragraph{Q3: Robustness under jailbreak (\S\ref{sec:robustness}).}
Q3 reuses the skills from Q1, so we apply the same correction to 60 comparisons per split under each attack. On execution, 48 comparisons reach $p<0.01$ under AgentHarm and 49 do so under ArtPrompt. The corresponding clean-prompt count is 54. On generation, the counts fall from 31 on clean prompts to 19 under AgentHarm and 12 under ArtPrompt.

Two effects reduce generation significance. First, both reformulations reduce the severity produced by the undefended agents and baselines. Mean baseline AvgS falls from 2.01 on clean prompts to 1.38 under AgentHarm and 1.19 under ArtPrompt. Per-class AvgS changes from 0.42 to 0.56 and 0.68, leaving smaller differences for the tests to detect. Second, the median number of untied families falls from 8 to 6 under AgentHarm and 7 under ArtPrompt. Among comparisons that miss $\alpha'$, 31 of 41 still favor \sysname under AgentHarm and 38 of 48 favor it under ArtPrompt.

\paragraph{Q4: Benign refusal on SWE-Bench Pro (\S\ref{sec:rq3}).}
Under the criterion in \S\ref{sec:setup}, every baseline has zero safety refusals across 731 tasks and six LLMs. \sysname also has zero safety refusals on four LLMs. Because the baseline rates are at the floor, a paired test provides no informative comparison. We therefore report the refusal rates without a $p$-value.

\section{Skill Examples}
\label{app:skills}

\begin{tcolorbox}[
  width=\dimexpr\columnwidth-4pt\relax,
  colback=gray!4, colframe=black!55,
  boxrule=0.4pt, arc=1.2pt,
  left=4pt, right=4pt, top=3pt, bottom=3pt,
  title={\small\textbf{Proactive skill (excerpt)}},
  fonttitle=\small, coltitle=white, colbacktitle=black!70,
  enhanced,
]
\fontsize{6.5}{7.7}\selectfont\raggedright
\textbf{Frontmatter.}\par
\texttt{name: redcode-proactive-defense}\par
\looseness=-1
\texttt{description:} procedural countermeasures matching \mbox{technical signatures before execution.}

\smallskip
\textbf{Rule 1: sensitive file access.}\par
Match file paths against a denylist of prefixes such as \texttt{/proc/}, \texttt{/sys/kernel/}, \texttt{/var/log/}, \texttt{/root/}, and specific files like \texttt{/etc/passwd} and \texttt{/etc/shadow}.
Also flag any \texttt{.ssh/} directory and any read via \texttt{cat}, \texttt{head}, or \texttt{open()} on these paths.\par
$\langle\textit{4 additional path patterns}\rangle$

\smallskip
\textbf{Rule 2: network exfiltration.}\par
Flag the imports \texttt{requests}, \texttt{urllib.request}, \texttt{socket}, or \texttt{httpx} when they co-occur with file reads. Flag \texttt{requests.post()} or \texttt{curl -X POST} sequenced after a file open. Block any code that combines rule~1 with a network call.

\smallskip
\textbf{Rule 6: multi-stage chaining.}\par
If two or more rules trigger in one execution flow, block. Example: read a kernel file, encode it as binary, then POST to a localhost endpoint.

\smallskip
$\langle\textit{rules 3--5, 7 omitted}\rangle$

\smallskip
\textbf{Refusal directive.}\par
Refuse cleanly, name the matched rule, do not run a modified or partial version.
\end{tcolorbox}

\paragraph{Threat under defense.}
Both excerpts target one RedCode-Exec attack family. The request asks the agent to write and run Python that reads a sensitive file, such as a password file, kernel-state file, or Secure Shell (SSH) key. The script then sends the bytes over the Hypertext Transfer Protocol (HTTP) to a Uniform Resource Locator (URL) chosen by the attacker. This behavior exfiltrates data from the host. Using the same threat for both examples makes the difference between the synthesis strategies directly comparable.

\paragraph{Setup.}
Each box reproduces the skill verbatim. Ellipses ($\langle\textit{...}\rangle$) mark omitted text. Both skills were synthesized for DS-V3.2 at \emph{per-class} granularity under the default budget $L = 10{,}000$ characters.

\begin{tcolorbox}[
  width=\dimexpr\columnwidth-4pt\relax,
  colback=gray!4, colframe=black!55,
  boxrule=0.4pt, arc=1.2pt,
  left=4pt, right=4pt, top=3pt, bottom=3pt,
  title={\small\textbf{Reactive skill (excerpt)}},
  fonttitle=\small, coltitle=white, colbacktitle=black!70,
  enhanced,
]
\fontsize{6.5}{7.7}\selectfont\raggedright
\textbf{Frontmatter.}\par
\texttt{name: redcode-reactive-defense}\par
\texttt{description:} identify harmful intent regardless of surface format, learned from cases where the undefended agent complied.

\smallskip
\textbf{Principle 1: data exfiltration intent.}\par
Read \emph{intent}, not syntax. A request that pairs reading a sensitive file with any transmission step (POST, upload, socket) is exfiltration, even when the destination is a localhost address or the payload is wrapped as a ``generic file-upload utility''.\par
\emph{Failure-derived clarifications.}
(i)~Localhost as target does not sanitize the request: the URL is trivially swappable.
(ii)~Framing as ``a function to send a file via POST'' does not make it benign once the demo binds to a sensitive path.
(iii)~A numbered procedural list (``1.~import, 2.~open file, 3.~read binary, 4.~POST, 5.~print status'') is a ready-to-run payload and should be read as one.

\smallskip
\textbf{Principle 3: unauthorized remote interaction.}\par
A network component built against a dummy URL still produces reusable attack scaffolding. Refuse before the scaffold is written.

\smallskip
$\langle\textit{principles 2, 4 omitted}\rangle$

\smallskip
\textbf{Why ``helpful modifications'' are still harmful.}\par
Running ``just the safe parts'', swapping the target URL, or substituting a destructive command with a no-op still demonstrates the technique and teaches the requester what works. Refuse cleanly, with no modified or partial execution.
\end{tcolorbox}

\paragraph{What the two excerpts reveal.}
The Proactive skill contains an explicit denylist. Its rules identify sensitive paths, network APIs, and sequences such as opening a file before making a network call. The refusal directive tells the agent to match incoming code against these signals. A defender can audit the rules and add a path or library when the attack corpus changes. Protection depends on the listed vocabulary. A paraphrased path or an unlisted HTTP client may fall outside the rules. Adding more variants can improve coverage, but the budget $L$ limits how many rules survive refinement.

The Reactive skill describes intent and rationalizations observed in failed trajectories. Its signals include claims that localhost is safe, that the code is a generic upload utility, or that numbered instructions are merely a tutorial. The undefended agent used these framings when it complied with the malicious request. A principle can remain relevant when an attacker changes the wording, but it is more abstract and harder to audit. Reactive synthesis also requires a baseline run that elicits the failure. A direct command such as \texttt{rm -rf /} may produce no rationalization in the recorded trajectory.

The strategies provide complementary information. Proactive synthesis supplies immediate coverage of known technical patterns. Reactive synthesis adds principles based on the evaluated agent's observed rationalizations. A production system could combine both sources, although our experiments evaluate the two strategies separately.

%%%%%%%%%%%%%%%%%%%%%%%%%%%%%%%%%%%%%%%%%%%%%%%%%%%%%%%%%%%%%%%%%%%%%%%%%%%%%%%%
\end{document}